\documentclass{ametsocV6.1_nolinenumbers}

\bibpunct{(}{)}{;}{a}{}{,}

\newcommand{\dd}[2]{{\frac{d #1}{d #2}}}
\newcommand{\DD}[2]{{\frac{D #1}{D #2}}}
\newcommand{\ddn}[3]{{\frac{d^{#3} #1}{d #2^{#3}}}}
\newcommand{\pp}[2]{{\frac{\partial #1}{\partial #2}}}

\newcommand{\eps}{{\varepsilon}}
\newcommand{\vu}{\mathbf{u}}
\newcommand{\vs}{\mathbf{x}}

\newcommand{\eq}[1]{(\ref{#1})}
\newcommand{\vvb}{\mathbf{v}}
\newcommand{\Order}[1]{{O} \! \left( #1 \right)}

\usepackage {ulem}

 \newcounter{sdcommentno}
\newcommand{\revision}[1]{\leavevmode{#1}}

\title{Interactions between gravity waves and cirrus clouds: asymptotic modeling of wave induced ice nucleation}

\authors{Stamen I.~Dolaptchiev, \aff{a}\correspondingauthor{Stamen
    Dolaptchiev, dolaptchiev@iau.uni-frankfurt.de}
  Peter Spichtinger, \aff{b}
  Manuel Baumgartner,\aff{b}
  and Ulrich Achatz\aff{a} } 

\affiliation{\aff{a}{Institut f\"ur Atmosph\"are und Umwelt,
    Goethe-Universit\"at Frankfurt, Frankfurt am Main, Germany}\\
  \aff{b}{Johannes Gutenberg-Universit\"at Mainz, Mainz, Germany}
  }

  \abstract{We present an asymptotic approach for the systematic
    investigation of the effect of gravity waves (GW) on ice clouds
    formed through homogeneous nucleation. In particular, we consider
    high- and mid-frequency GW in the tropopause region driving the
    formation of ice clouds, modeled with a double-moment bulk ice
    microphysics scheme. The asymptotic approach allows for
    identifying reduced equations for self-consistent description of
    the ice dynamics forced by GW including the effects of diffusional
    growth and nucleation of ice crystals. Further, corresponding
    analytical solutions for a monochromatic GW are derived under a
    single-parcel approximation. It is demonstrated that the
    asymptotic solutions capture the dynamics of the full ice model
    and provide a simple expression for the nucleated number of ice
    crystals. The present approach is extended to allow for
    superposition of GW, as well as, for variable mean mass in the ice
    crystal distribution. Implications of the results for an improved
    representation of GW variability in cirrus parameterizations are
    discussed.}

\begin{document}

\maketitle

\section{Introduction}
Clouds consisting exclusively of ice particles, so-called cirrus
clouds, account for roughly one third of the total cloud cover
\citep[e.g.,][]{gasparini_cirrus_2018}, yet their net radiative effect
is still one major source of uncertainty in the climate system. Since
albedo effect and greenhouse effect are of the same order of magnitude
for those clouds, microphysical details of ice crystals \citep[as,
e.g., shape or size, see][]{zhang_etal1999, kramer_microphysics_2020}
may determine the net radiative effect.  The microphysical properties,
however, are strongly influenced by a complex interplay of nuclei
composition, micro-scale cloud processes and multi-scale interactions
with the surrounding atmosphere. All those components are poorly
understood and, if at all, only crudely represented in climate models.

Cirrus clouds can be subdivided into liquid origin and in situ cirrus
\citep[e.g.][]{kramer_microphysics_2016}. The former class describes
clouds originating from cloud droplets, which freeze in upward
motions, e.g., in mesoscale convective outflow or warm conveyor
belts. In contrast, the ice crystals of in situ cirrus are formed
without any pre-existing cloud droplets: either by homogeneous
freezing of aqueous solution droplets \citep[short: homogeneous
nucleation, see, e.g., ][]{koop_water_2000, baumgartner_new_2022}, or
by heterogeneous nucleation
\citep[e.g.][]{pruppacher_microphysics_2010, hoose_moehler2012,
  baumgartner_new_2022} initiated by solid aerosol particles.

Observational studies indicate that cirrus properties and life cycle
can be crucially affected by gravity wave (GW) dynamics
\citep[e.g.][]{karcher_roles_2003, kim_ubiquitous_2016,
  bramberger_first_2022}. GW are generated to a large fraction in the
troposphere and often propagate over considerable horizontal and
vertical distances before breaking. During their propagation those
waves can generate substantial oscillations in the atmospheric fields
and the GW drag, exerted in the breaking region, alters the mean
atmospheric state. Because of the importance of small-scale GW
dynamics some cirrus studies explicitly resolve the GW using LES
models \citep[e.g.][]{joos_orographic_2009,
  kienast-sjogren_formulation_2013}, or detailed parcel models
\citep[e.g.][]{haag_kaercher2004, jensen_pfister2004,
  spichtinger_kraemer2013}.

\revision{There are several classes of schemes for modeling the
  influence of GW on ice clouds in climate models. In all schemes, a
  subgrid scale GW vertical velocity is diagnosed and then directly
  used in the cirrus cloud scheme. One should keep in mind that
  climate models diagnose the number concentration of ice crystals in
  a (homogeneous) nucleation from the vertical velocity \citep[see,
  e.g.,][]{karcher_parameterization_2002, ren_mackenzie2005,
    wang_cirrus_2010}. First, there are schemes without any physical
  constraint on GW. These schemes mostly rely on the use of a
  turbulent kinetic energy (TKE) scheme in the upper troposphere. This
  approach is per se questionable since most TKE schemes were
  developed for parameterizing turbulence in the planetary boundary
  layer. The TKE approach is known to produce quite high vertical
  velocities \citep[see, e.g.,][]{joos_orographic_2008, zhou_etal2016}
  and the pattern of enhanced vertical velocities usually do not agree
  with regions of enhanced GW activity. Second, there is a general
  approach using distributions of temperature fluctuations constructed
  from measurements. From these subgrid scale vertical velocities are
  derived \citep{kaercher_burkhardt2008, wang_cirrus_2010,
    podglajen_lagrangian_2016, karcher_stochastic_2019-1}. So far such
  approaches do not take into account GW directly but the GW signal is
  masked by other effects and there is no direct link to the GW
  sources, such as mountains, convection, spontaneous imbalance or
  others. One should also keep in mind that the measurements are
  sparse and they are largely extrapolated into other regions without
  any measurements. Further, it is not apriori clear that the
  underlying statistical description of temperature fluctuations will
  remain unaltered under climate change. Finally, there are some
  attempts of diagnosing the GW vertical velocity using linear theory
  for mountain waves \citep{dean_parameterisation_2007,
    joos_orographic_2008}. In both schemes there is a direct
  connection between the source of GW and the cloud scheme, which is a
  clear advantage in comparison to schemes relying on statistical
  information. However, no other sources than mountain waves are
  represented up to now.}


Most of the current GW parameterizations in climate models rely on the
single-column, steady-state approximation. Under this assumption GW
propagate only in the vertical and instantly fast up to the breaking
altitude, where they deposit energy and momentum. The limitations of
steady-state parameterizations were demonstrated in the study by
\citet{boloni_interaction_2016}, where a transient approach was
proposed. The new transient parameterization was implemented by
\cite{boloni_toward_2021} in the weather-forecast and climate model
ICON and \revision{\cite{kim_toward_2021} showed that the resulting
  intermittency patterns of convectively generated GW are similar to
  observations. Thus, developing a cirrus scheme to be coupled to a
  transient GW parameterization is a promising route for more
  realistic representation of ice clouds in climate models. Such
  development} requires the systematic identification of the dominant
interaction processes between GW and cirrus and their self-consistent
description. \cite{baumgartner_homogeneous_2019} , hereafter BS19,
utilized a matched-asymptotic approach \revision{\citep[e.g., see
  ][for an introduction to asymptotics]{holmes}} for studying
homogeneous nucleation due to constant updraft velocities. The
resulting parameterization successfully reproduces the results of the
classical scheme of \cite{karcher_parameterization_2002}. Encouraged
by the results of BS19, we extend their asymptotic approach to allow
for GW dynamics. We construct a self-consistent simplified model for
GW-cirrus interactions and corresponding asymptotic solutions
applicable for diagnosing ice crystal numbers in nucleation events
forced by passing GW.  \revision{An application of our analytical
  approach would be a direct coupling of the transient GW
  parameterization \citep[][]{boloni_toward_2021, kim_toward_2021} to
  our analytical model. The GW parameterization will provide
  information about the wave amplitudes, frequencies and wavenumbers,
  which can directly be used for our approach. The detailed
  information on the wave spectrum allows to predict the ice crystal
  number concentration more realistically than simple diagnostic
  relations in large-scale models
  \citep[e.g.][]{karcher_parameterization_2002}, which are based on
  constant vertical updraft motion.}

This paper is organized as follows: The asymptotic representation of
the GW and ice microphysics can be found in
section~\ref{asymptotic_approach}.  In section~\ref{derivation} we
derive the reduced equations describing the GW and the cirrus
dynamics. In addition, asymptotic solutions are constructed, modeling
the nucleation, as well as, the pre- and post-nucleation
dynamics. Numerical simulations of the full ice physics model and
validation of the asymptotic solutions can be found in
section~\ref{numerical_experiments}. In section \ref{sec_var_mass} the
present approach is extended to take into account variations in the
mean mass of the ice crystal distribution. \revision{In
  section~\ref{superposition} the asymptotic solution is extended to
  the case of multiple GW driving the ice physics.} Concluding
discussions are summarized in section~\ref{conclusions}.

\section{Asymptotic approach for studying GW-cirrus interactions}
\label{asymptotic_approach}

\subsection{Gravity wave dynamics: governing equations  and scalings}
We start with the equations governing a compressible flow on a
$f$-plane \citep[e.g.][]{durran_improving_1989}, without diabatic and
frictional sources
\begin{align}
\label{dim_dudt}
  \DD{\vu}{t} + f \mathbf{e}_z \times \vu  &= 
- c_p \theta \nabla_h \pi\, , \\
\DD{w}{t} &= - c_p \theta \pp{\pi}{z} - g\, ,\\
\DD{\theta}{t} &= 0\, ,\\
\label{dim_dpdt}
  \DD{\pi}{t} + \frac{R \pi}{c_v} \nabla \cdot \vvb &= 0\, .
\end{align}
Here, the total wind vector $\vvb$ is separated into a horizontal,
$\vu$, and vertical, $w$, component, $g$ denotes the gravitational
acceleration, $f$ the Coriolis parameter and $\DD{}{t}$ the material
derivative. In addition, $c_p$ and $c_v$ are the specific heat
capacities of dry air at constant pressure and volume, respectively,
and the ideal gas constant is given by $R=c_p -c_v$. Further, $\theta$
denotes the potential temperature based on constant $c_p$
\citep{baumgartner_reappraising_2020}. The Exner pressure $\pi$ is
related to the pressure $p$ by $\pi = (p/p_{00})^{R/c_p}$, where
$p_{00}$ is some reference pressure. The ideal gas law $p/\rho = R T$
is assumed to be valid, where $\rho$ is density and $T$
temperature. We consider a hydrostatically balanced reference
atmosphere at rest with pressure scale height, $H_p$, and potential
temperature scale height, $H_\theta$, which depend on altitude and are
defined by
\begin{align}
  H_p^{-1} = \left|\frac{1}{\bar p}\dd{\bar p}{z}\right| =
  \frac{g}{R\bar T}\, ,\qquad 
  H^{-1}_\theta =  \frac{1}{\bar \theta}\dd{\bar \theta}{z}
  =  \frac{1}{\bar T}\left(\dd{\bar T}{z} + \frac{g}{c_p}\right)\, . 
\end{align}
In the last equations variables with an overbar refer to the
reference atmospheric fields.

Within the framework of multiscale asymptotics, we have to specify a
distinguished limit in order to define the regimes we are interested
in, this is carried out in the following way:

First, we allow for weak and moderately strong stratification
appropriate for the dynamics in the upper troposphere and lower
stratosphere. Following \cite{achatz_interaction_2017}, the different
stratifications can be expressed using the ratio
\begin{align}
  \label{hp_to_ht}
 \frac{H_p}{H_\theta} = \eps^\alpha\, , \quad \text{ where }
 \eps=\Order{10^{-1}}\quad \alpha=0,1\, . 
\end{align}
In the equation above $\alpha=0$ corresponds to the strong and
$\alpha=1$ to the weak stratification case. 

Second, we specify the frequency regime of the GWs. 
In the analyses here the high- as well as mid-frequency
GW are considered. Using the Brunt-Vaisala frequency
$N = \sqrt{\frac{g}{\bar \theta} \dd{\bar \theta}{z}}$ the reference
GW time scale, $T_w$, is defined as
    \begin{align*}
      T_w = \frac{1}{\eps^{\beta}N} \, , \quad \beta=0,1   
    \end{align*}
    where $\beta = 0$ characterizes the high-frequency and $\beta=1$
    the mid-frequency GW time scale. The corresponding GW period,
    $P_w$, is given by $P_w = 2 \pi T_w$. For the vertical length scale
    of the GW, $H_w$, we assume \citep{achatz_gravity_2010,
      achatz_interaction_2017}
\begin{align}
  H_w = \eps H_p\, .
\end{align}
The appropriate horizontal length scale, $L_w$, is estimated using the
inertial GW dispersion relation
\citep[e.g.,][]{achatz_atmospheric_2022} 
\begin{align}
  \label{IGW_despersion_relation}
  \hat \omega^2 = \frac{f^2 m^2 + N^2 k_h^2}{k_h^2 + m^2}\, ,
\end{align}
with $\hat \omega$ the intrinsic GW frequency, $m$ the vertical
wavenumber and $k_h$ the magnitude of the horizontal wave vector. By
setting $f/N = \Order{\eps^{\frac{5-\alpha}{2}}}, \hat \omega =
1/T_w$, $m=1/H_w$ and $k_h = 1/L_w$ one obtains the estimate
\begin{align}
 L_w = \eps^{-\beta} H_w\, . 
\end{align}
Since the aspect ratio $H_w/L_w$ defines the anisotropy of the GW, the
high- and mid-frequency GW correspond to isotropic and moderately
anisotropic waves, respectively. The reference quantity for the
horizontal \revision{wave} velocity scale, $U$, and for the vertical
\revision{wave} velocity scale, $W$, are estimated using the advection
velocities
\begin{align}
U = \frac{L_w}{T_w} \, ,\qquad W = \frac{H_w}{T_w} \, .
\end{align}
As shown in \cite{achatz_interaction_2017} the above expressions are
consistent with the polarization relations for GW if the mean flow
entering the Doppler term is not larger than $U$. By introducing a
reference temperature $T_{00}$ such that
\begin{align}
\bar \theta = \Order{T_{00}}
\end{align}
and using $\bar T = \Order{T_{00}}$, one arrives at the expressions
\begin{align}
  H_p = \Order{\frac{R T_{00}}{g}}\, ,\quad
  U = \Order{\eps^{\frac{2+\alpha}{2}} \sqrt{R T_{00}}}, \quad
  W =  \Order{\eps^{\frac{2+\alpha+2\beta}{2}} \sqrt{R T_{00}}} \, .
\end{align}
Using the above scaling for $W$ we allow for strong and moderate
vertical velocities: these are of the order of $U$ in the case of
high-frequency GW and of the order of $\eps U$ for the mid-frequency
GW. The scalings presented in this section imply the following
distinguished limit for the Mach, Froude and Rossby numbers
\begin{align}
  \mathrm{Ma} = \frac{U}{\sqrt{R T_{00}}} \sim \eps^{\frac{2+\alpha}{2}}\, ,\quad
  \mathrm{Fr} = \frac{U}{N H_p} \sim \eps\, , \quad
  \mathrm{Ro} = \frac{U}{f L_w} \sim \eps^{\frac{2\beta + \alpha -5}{2}}\, .
\end{align}
The magnitude of the buoyancy GW fluctuations, $B_w$, is set to the
one associated with GW close to breaking due to static instability
\citep{achatz_gravity_2010}, namely
\begin{align}
  \label{scaling_b}
  B_w = N^2 H_w = \eps^{\alpha+1} g\, .
\end{align}
From the buoyancy definition $b = g \frac{\theta - \bar \theta}{\bar \theta}$
one obtains for the magnitude of GW potential temperature fluctuations   
\begin{align}
  \label{scaling_theta}
 \Theta_w = \eps^{\alpha+1} T_{00} \, .
\end{align}
Finally, as shown in \cite{achatz_interaction_2017} from the
polarization GW relations, the GW \revision{Exner} pressure fluctuations scale as  
\begin{align}
\label{scaling_pi}
  \Pi_w \sim \frac{i}{m}\frac{\hat \omega^2 - N^2}{N^2}\frac{B_w}{c_p
    \bar \theta} = \Order{\eps^{2+\alpha}}\, . 
\end{align}
Nondimensionalizing the governing equations
\eq{dim_dudt}-\eq{dim_dpdt} with the reference quantities from
Table~\ref{gw_scaling} and replacing
\begin{align}
  (x, y, z, t, \nabla_h) &
  \rightarrow (L_w x_w, L_w y_w, H_w z_w, T_w t_w, L^{-1}_w \nabla_h) \\
  (\mathbf{u}, w, \theta, \pi, T, p) &
  \rightarrow (U \mathbf{u}, W w, T_{00}\theta, \pi, T_{00}T, p_{00}p) \\
   (f, g)&\rightarrow (\eps^{5/2} \frac{g}{\sqrt{R T_{00}}} f, g)
\end{align}
yields
\begin{align}
  \label{nondim_dudt}
  \eps^{2+\alpha}\DD{\vu}{t_w} + \eps^{\frac{9+\alpha}{2}-\beta} \mathbf{e}_z
  f \times \vu  &=  
- \frac{c_p \theta}{R} \nabla_h \pi\, , \\
  \eps^{2+\alpha+2\beta} \DD{w}{t_w}
&= -\frac{c_p \theta}{R} \pp{\pi}{z_w} - \eps\, , \\
\label{nond_dot_theta}
  \DD{\theta}{t_w} &= 0\, ,\\
  \label{nondim_dpdt}
  \DD{\pi}{t_w} + \frac{R \pi}{c_v} \nabla \cdot \vvb &= 0\, .
\end{align}
\begin{table}
  \centering
  \begin{tabular}[hb!]{c | c }
\topline
    ref. quantity & value \\
    \midline
    $H_w$ & $\eps H_p=\eps \frac{RT_{00}}{g}$ \\
       $T_w$ & $\eps^{-\beta} N^{-1}$  \\
    $L_w$ & $\eps^{1-\beta} H_p = \eps^{1-\beta}\frac{RT_{00}}{g} $ \\
    $U$ & $\eps^{\frac{2+\alpha}{2}}\sqrt{RT_{00}}$ \\
$W$ & $\eps^\beta U=\eps^{\frac{2+\alpha + 2\beta}{2}}\sqrt{RT_{00}} $ \\
$\Theta_w$ & $\eps^{1+\alpha}T_{00} $ \\ 
    $\Pi_w$ & $\eps^{2+\alpha}$ \\
    $f$ & $\eps^{\frac{5-\alpha}{2}}N$ \\
    \botline
  \end{tabular}
  \caption{Reference quantities for high-frequency, $\beta=0$, and
    mid-frequency, $\beta=1$, gravity wave scaling. Here we choose
    $T_{00}=210$ K and $N=10^{-2}$ s$^{-1}$ for the troposphere and
    $N=2\times 10^{-2}$ s$^{-1}$ for the tropopause
    region. \revision{Note, however, that regimes with other values of
      $T_{00}$ and $N$ can be considered as well.}}
\label{gw_scaling}
\end{table}
\subsection{Ice microphysics:  governing equations and scalings}
\label{sec:ice_eq}
The cirrus clouds are described by a double-moment bulk microphysics
scheme assuming an unimodal ice mass distribution function. The scheme
is the same as the one from BS19, except that the sedimentational
sinks are included here. A more detailed description of the ice model
can be found in \cite{spichtinger_modelling_2009} and
\cite{spreitzer_subvisible_2017}, in this section we only briefly
refer to some key properties. As in BS19 we assume spherical shape of
ice crystals, which leads to a simpler description of the cloud
processes.

The equations governing the ice crystal number
concentration $n$ (number of ice crystals per mass dry air, unit
[kg$^{-1}$]), ice mixing ratio $q$ (mass of ice per mass dry air, unit
[kg/kg]) and vapor mixing ratio $q_v$ (mass of water vapor per mass
dry air, unit [kg/kg]) read
\begin{align}
  \label{dot_n_scheme}
 \DD{n}{t} &= \mathrm{Nuc}_n + \mathrm{Sed}_n  \\
  \label{dot_q_scheme}
  \DD{q}{t} &= \mathrm{Dep} + \mathrm{Nuc}_q + \mathrm{Sed}_q  \\
  \label{dot_v_scheme}
  \DD{q_v}{t} &= -\mathrm{Dep} - \mathrm{Nuc}_q\, ,
\end{align}
where $\mathrm{Dep}$ describes the ice crystal growth due to the deposition of
water vapor, $\mathrm{Nuc}$ the generation of new ice crystals through
homogeneous nucleation and $\mathrm{Sed}$ the sedimentation of ice
crystals under the effect of gravity. The latter sedimentational
processes are modeled as
\begin{align}
 \mathrm{Sed}_{n} &= \frac{1}{\rho} \pp{}{z} \rho v_{n} n\, ,\\
 \mathrm{Sed}_{q} &= \frac{1}{\rho} \pp{}{z} \rho v_{q} q\, ,
\end{align}
where we assume spatially independent sedimentation velocities
$v_{n,q} = c_{n,q} m_{\text{ref}}^{2/3}$ with constants $c_n = 5.8 \times
10^5\, ,$ $c_q = 1.2 \times 10^6$ in units of m s$^{-1}$ kg$^{-2/3}$ and a
reference mass $m_{\text{ref}}$; this simplification is sufficient for
estimating typical values of the sedimentation terms in the following
asymptotic analysis.

In \eq{dot_q_scheme}, \eq{dot_v_scheme} the
deposition term $\mathrm{Dep}$, also referred to as diffusional growth term,
can be parameterized as
\begin{align}
  \label{dep}
  \mathrm{Dep} = C_0 \bar m^{1/3} \frac{p_{si}}{p} (S - 1) T n
\end{align}
with $C_0 = 4.3\, 10^{-8}$ kg$^{2/3}$ s$^{-1}$K$^{-1}$, $p_{si}$ the saturation
pressure over flat ice surface, $\bar m=q/n$ the mean ice-particle
mass and $S$ the saturation ratio with respect to ice. The latter is
defined as
$$S = \frac{p_v}{p_{si}}\, ,$$
where $p_v$ is the water vapor pressure. Using the
definition of $q_v=m_v/m_d=\rho_v/\rho_d$ and the ideal gas law to
express the dry air pressure, $p_d = R T \rho_d$, and the water vapor
pressure, $p_v = R_v T \rho_v$, yields for the saturation ratio
\begin{align}
S = \frac{q_v p_d}{ \eps_0 p_{si}} \approx \frac{q_v p}{ \eps_0 p_{si}} \, ,
\end{align}
where $\eps_0=R/R_v$ and the water vapor gas constant is given by
$R_v=461$ J kg$^{-1}$ K$^{-1}$. The saturation pressure $p_{si}$ is
highly dependent on temperature and satisfies the approximate
Clausius-Clapeyron equation
\begin{align*}
      \dd{p_{si}}{T} = \frac{L_i}{R_v T^2}p_{si}
\end{align*}
with $L_i=2.8 \times 10^6$ J kg$^{-1}$. In eq. \eq{dot_n_scheme} the homogeneous
nucleation rate of ice crystals is modeled as
\begin{align}
  \mathrm{Nuc}_n = J \exp\left(B(S -S_c) \right)
\end{align}
\citep[following BS19 and][]{spichtinger_impact_2022},
where $S_c$ is some critical saturation ratio $S_c(T) \approx 1.5$,
$J = 4.9\, 10^4$ kg$^{-1}$ s$^{-1}$ and $B = 337$ (see BS19 for further details
and the estimation of these values). The nucleation term in the
equation for ice mixing ratio \eq{dot_q_scheme} is given by
\begin{align}
  \mathrm{Nuc}_q = \hat m_0 \mathrm{Nuc}_n\, ,
\end{align}
where the reference mass $\hat m_0 = 10^{-16}$ kg $\ll \bar m$ is
used, which represents a typical mass of newly nucleated ice
crystals. A summary of the ice physics scheme parameters can be found
in Tab.~\ref{tab:ref2}. Next, characteristic numbers for the ice
physics variables are chosen.  Those values should describe a typical
cirrus cloud in the upper troposphere/ lower stratosphere region
formed due to homogeneous nucleation. The characteristic values for
the number concentration, vapor mixing ratio and ice mixing ratio are
denoted by $n_c$, $q_{vc}$ and $q_c$, respectively. The estimate
$m_{\text{ref}} \sim \bar m \sim m_c$ is used, where $m_c$ denotes some mean
ice crystal mass satisfying $q_c = n_c m_c$. All reference values can
be found in Tab.~\ref{tab:ref1}, they agree with the characteristic
values used in BS19.
\begin{table}
  \centering
  \begin{tabular}[h]{c | c }
\topline
    quantity & value \\
    \midline
    $J$ & $4.9\times 10^4$ kg$^{-1}$ s$^{-1}$  \\
    $B$ & $337$ \\
    $S_c$ & $1.5$ \\
    $C_0$  & $4.3\times 10^{-8}$ kg$^{2/3}$ s$^{-1}$ K$^{-1}$ \\
    $\eps_0$ & $0.62$ \\
    $c_n$  & $5.8\times 10^5$ m s$^{-1}$ kg$^{-2/3}$  \\
    $c_q$ & $12\times 10^5$ m s$^{-1}$ kg$^{-2/3}$ \\
    $\hat m_0$ &$10^{-16}$ kg  \\
    $L_i$ & $2.8\times 10^6$ J kg$^{-1}$  \\
    $R_v$ & $461$ J kg$^{-1}$ K$^{-1}$  \\
    \botline
  \end{tabular}
  \caption{Parameters of the ice physics scheme}
\label{tab:ref2}
\end{table}

\begin{table}
  \centering
  \begin{tabular}[h]{c | c}
\topline
    ref. quantity & value \\
\midline
    $n_{c}$ &  $2\times  10^6\ \text{kg}^{-1}$ \\
    $q_{v,c}=\frac{\epsilon_0 p_{si,c}}{p_{00}}$ & $2\times 10^{-5}$ \\
    $m_{c}=m_{\text{ref}}=\bar m$ &  $10^{-12}\ \text{kg}$ \\
    $q_c=m_c n_c$  & $2\times 10^{-6}$ \\ 
    $T_d = \left(C_0 m_c^{1/3}
    \frac{p_{si,c}}{p_{00}\, q_{v,c}} T_{00} n_c\right)^{-1}$
                  & $\revision{340}$ s\\
    $H_c$  & $600$ m \\
     $p_{00}$ & $300$ hPa \\
     $T_{00}$ & $210$ K \\
     $\rho_{00}$ & $0.5\ \text{kg m}^{-3}$ \\
     $p_{si,c}(T_{00})$ & $1$ Pa  \\
    \botline
  \end{tabular}
  \caption{Reference quantities used for non-dimensionalization of the
    ice physics scheme} 
\label{tab:ref1}
\end{table}
Next, \revision{from \eq{dep} a characteristic time scale on which the
diffusional growth term acts can be introduced \cite[see also][]{korolev_supersaturation_2003,kramer_ice_2009}}, it is defined as
\begin{align}
  T_d = \left(C_0 m_c^{1/3} \frac{p_{si,c}}{p_{00}\, q_{v,c}} T_{00}
  n_c\right)^{-1} \sim \revision{340}\ \text{s}\, ,
\end{align}
if the reference saturation pressure over ice $p_{si,c}=1$ Pa is used
(see Tab.~\ref{tab:ref1} for all other values). As can easily be shown,
the estimate above for $T_d$ allows for deviations of about $10$~K
from the reference temperature $T_{00}=210$~K.  The characteristic
vertical scale of the cirrus cloud is set to
\begin{align}
 H_c \sim H_w\, . 
\end{align}
This corresponds to $H_c= 600$ m for $H_w$ from Tab.~\ref{gw_scaling}.
Finally, the ice physics scheme is nondimensionalized using the
reference quantities and all arising nondimensional numbers are
expressed in terms of $\eps$ (distinguished limit), as summarized in
Tab.~\ref{tab:dist_limit}. Applying the replacements
\begin{align}
  (z, t) &
  \rightarrow (H_w z_w, T_d t_d) \\
  (n,\, q,\, q_{v},\, p_{si},\, \rho) &
  \rightarrow (n_c n,\, q_c q,\, q_{v,c}q_v,\ p_{si,c} p_{si},\ \rho_{00}\rho)
\end{align}
one yields the following nondimensional equations
\begin{align}
  \label{dot_n_non}
  \DD{n}{t_d} &= \frac{J^*}{\eps} \exp \left(\frac{B^*}{\eps^2} (S
  -S_c) \right) +  
  \eps^{2}\frac{S^*_n}{\rho} \pp{}{z_w} \rho  n  \\
  \DD{q}{t_d} &= \frac{1}{\eps}D^* \frac{p_{si}}{p}(S-1) T n + 
  \eps^{3}J^* \exp \left(\frac{B^*}{\eps^2} (S -S_c) \right)
  +\eps^{2}\frac{S^*_q}{\rho} \pp{}{z_w}  \rho q     \\
  \label{q_v}
  \DD{q_v}{t_d} &= - D^* \frac{p_{si}}{p}(S-1) T n 
  - \eps^{4}J^* \exp \left(\frac{B^*}{\eps^2} (S -S_c) \right)\, ,
\end{align}
where an asterisk denotes an order one constant.
\begin{table}
  \centering
  \begin{tabular}[h]{c | c | c }
\topline
    quantity & value & distinguished limit\\
    \midline
    $B$ & $337$ & $\frac{B^*}{\eps^2}$ \\
    $ \frac{C_0 m_c^{1/3} T_{00} n_{c} T_w}{\eps_0} $ & $1.456 $ & $D^*$ \\
    $ \frac{C_0 m_c^{1/3} T_{00} n_{c} T_w q_{v,c}}{\eps_0 q_c}$ & $14.56$ & $\eps^{-1} D^*$  \\
    $\frac{c_n m_c^{2/3} T_w}{H_c}$ & $4.8\times 10^{-3}$ & $\revision{\eps^2} S^*_n$\\
    $\frac{c_q m_c^{2/3} T_w}{H_c}$ & $.01$ & $\eps^2 S^*_q$ \\
    $\frac{J T_w}{n_c}$ & $12.5$ & $\frac{J^*}{\eps}$\\
    $\frac{\hat m_0 J T_w}{q_c} $  & $1.2\times 10^{-3}$ & $\eps^3 J^*$ \\
    $\frac{\hat m_0 J T_w}{q_{v,c}} $  & $1.2\times 10^{-4}$ & $\eps^4 J^*$ \\
    $\frac{L_i}{R_v T_{00}}$ & $29$ & $\frac{L^*}{\eps}$\\
    \botline
  \end{tabular}
  \caption{Distinguished limits for the non-dimensional numbers in the
    ice scheme. In the right-most column a star denotes an order one
    constant. For the nondimensionalization a time scale $T_w$, with
    $T_w \sim T_d$, was used.}
\label{tab:dist_limit}
\end{table}

One will also make use of the non-dimensional form of the
Clausius-Clapeyron equation, which reads
\begin{align}
\label{cla_cla}
  \dd{p_{si}}{T} = \frac{L^*}{\eps T^2}p_{si}\, .
\end{align}
\revision{Finally, as shown in Appendix~A the evolution
  equation for $q_v$ is rewritten in terms of the saturation ratio
  $S$. With this, the nondimensional system governing the ice dynamics
  reads}
\begin{align}
  \label{n}
  \DD{n}{t_d} &= \frac{J^*}{\eps} \exp \left(\frac{B^*}{\eps^2} (S
  -S_c) \right) +  
  \eps^{2}\frac{S^*_n}{\rho} \pp{}{z_w} \rho  n  \\
  \label{q}
  \DD{q}{t_d} &= \frac{1}{\eps}D^* \frac{p_{si}}{p}(S-1) T n + 
  \eps^{3}J^* \exp \left(\frac{B^*}{\eps^2} (S -S_c) \right)
  +\eps^{2}\frac{S^*_q}{\rho} \pp{}{z_w}  \rho q     \\
  \label{S}
  \DD{S}{t_d} &= - D^* (S-1) T n 
  -\frac{S}{\pi} \DD{\pi}{t_d}
  \left(\frac{L^*}{\eps T} -
  \frac{c_p}{R}\right)
  - \eps^{4}\frac{p}{p_{si}}J^* \exp \left(\frac{B^*}{\eps^2} (S -S_c)
  \right)\, . 
\end{align} 
\subsection{Asymptotic expansion}
The nondimensional coordinates $\vs_w,t_w$, entering in equations
\eq{nondim_dudt}-\eq{nondim_dpdt}, describe variations on the GW
spatial and temporal scales. In order to take into account variations
of the reference atmosphere on the large, i.e. synoptic,
vertical scale, $H_s$, with $H_s\sim H_p=H_w/\eps$, we introduce a compressed
coordinate defined as
\begin{align}
  z_s = \eps z_w\, .
\end{align}
We consider a wave field (denoted by a prime) superimposed on a
hydrostatically balanced reference atmosphere (denoted by a bar)
\begin{align}
  \theta &= \bar \theta(z_s) + \eps^{1+\alpha}\theta'(\mathbf{x}_w, t_w)\\
  \label{expansion_pi}
  \pi &= \bar \pi(z_s) + \eps^{2+\alpha}\pi'(\mathbf{x}_w, t_w)\\
  \vvb &= \vvb' (\mathbf{x}_w, t_w)
\end{align}
where we have used the scaling from \eq{scaling_theta}, \eq{scaling_pi}.
Next, in accordance with \eq{hp_to_ht} the potential temperature of the reference atmosphere is expanded as
\begin{align}
  \bar \theta(z_s) =
  \begin{cases}
     \bar \theta^{(0)}(z_s) + \Order{\eps} &\text{ if } \alpha = 0 \\
     1 + \eps \bar \theta^{(1)}(z_s) + \Order{\eps^{2}}
    &\text{ if } \alpha = 1 
  \end{cases}\, , 
\end{align}
whereas the corresponding Exner pressure expansion reads
\begin{align}
  \bar \pi(z_s) = \sum\limits_{j=0}^{1+\alpha}
  \eps^j \bar \pi^{(j)}(z_s) + \Order{\eps^{\alpha+2}} \, .
\end{align}
For the wave part we make a wave ansatz, e.g. for the
potential temperature field it reads
\begin{align}
  \theta'(\mathbf{x}_w, t_w) = \mathrm{Re} \Big\{
  \tilde \theta^{(1+\alpha)} \exp[i(\mathbf{k} \cdot \mathbf{x}_w - \omega t_w)]\Big\}
  + \Order{\eps}\, ,
\end{align}
with the wave amplitude $\tilde \theta^{(1+\alpha)}$, the wave vector
$\mathbf{k} = (k, l, m)^T$ and the frequency $\omega$.  Further,
consistent with the definitions of the potential temperature and Exner
pressure one obtains
\begin{align}
  \label{expansion_T_rho}
  (T, \rho) &= (\bar T, \bar \rho)(z_s) +  \eps^{1+\alpha} (T',\rho')(\mathbf{x}_w, t_w, z_s)\\
  \label{expansion_p}
  p &= \bar p(z_s) +  \eps^{2+\alpha}p'(\mathbf{x}_w, t_w, z_s)\, .
\end{align}
%
The following asymptotic expansion for the ice fields $\mathbf{\chi}=(n,q,S)$ is used
\begin{align}
  \label{expansion_ice}
 \mathbf{\chi} &= \mathbf{\chi}^{(0)} + \Order{\eps}\, ,
\end{align}
in addition $S_c=\Order{1}$ is assumed.
Integrating \eq{cla_cla} with the boundary condition $p_{si}(1)=1$ (or in dimensional form $p_{si}(T_{00})=p_{si,c}$ ), one obtains for the saturation pressure over ice
\begin{align}
\label{p_sat_ice}
  p_{si}(T) &= \exp\left\{\frac{L^*}{\eps}\left(1-\frac{1}{\bar T + \eps^{1+\alpha}T'}\right)\right\}\, .  
\end{align}
In the next section we will consider the ice physics at the reference height $z_{00}$, at this level we have 
\begin{align}
  \label{bar_pT_z00}
  \bar p^{(0)} = \bar T^{(0)} = \bar \rho^{(0)} = \bar \theta^{(0)} = \bar \pi^{(0)} = 1
\end{align}
and from \eq{p_sat_ice}: $p_{si}(z_{00})=\Order{1}$.
%
%

\subsection{Coupling of the GW and diffusion time scale}
\label{sec:coupling_gw_cirrus_time_scale}
We consider the following distinguished limit for the GW time scale, $T_w$,
and for the diffusion time scale, $T_d$,
\begin{align}
  \label{dist_limit_t}
  \frac{T_d}{T_w} = \Order{1}\, .
\end{align}
Since $T_d=\revision{340}$ s, the scaling above is valid for mid-frequency GW in
the troposphere and stratosphere, as well, for high-frequency GW in
the troposphere. In the case of mid-frequency GW one has $T_w\sim10^3$
s in the troposphere and $T_w \sim 500$ s in the tropopause region, if
the Doppler term in the GW dispersion relation is neglected. In the
case of high-frequency GW in the troposphere one has $T_w \sim 100$
s. Note, that the corresponding GW period, $P_w$, reads
$P_w = 2 \pi T_w$. \revision{For high-frequency GW in the tropopause
  region $T_w \sim 50$ s and the resulting scaling is discussed in
  Sec.~\ref{superposition}.  For low-frequency GW in the tropopause
  region $T_w \sim 5000$ s is more appropriate. In this case
  $T_d/T_w = \Order{\eps}$ leads to a weak amplitude GW forcing and
  the corresponding regime will be presented in an upcoming study.}
The condition \eq{dist_limit_t} together with \eq{expansion_pi}
yields the transformation
\begin{align}
  \label{transf_dot_pi}
  \DD{\pi}{t_d} = \left(\eps^{2+\alpha}\DD{\pi'}{t_w} + \eps w\dd{\bar \pi}{z_s} \right)
\end{align}
Eq.~\eq{dist_limit_t} implies that the coordinates $t_w$ and $t_d$
resolve variations on the same time scale, hence we may identify these
two and replace $t_d$ in the following by $t_w$.
\section{Reduced model of GW-cirrus interactions}
\label{derivation}
\subsection{GW dynamics}
The asymptotic analysis of equations \eq{nondim_dudt}-\eq{nondim_dpdt}
gives that the leading order fields satisfy the GW polarization
relations
\begin{align}
  \label{polarization}
  \left(\tilde u^{(0)}, \tilde v^{(0)}, \tilde w^{(0)},
  \tilde \pi^{(2+\alpha)} \right) =
  -\frac{im\omega}{k_h^2 \bar N^2}\tilde b^{(1+\alpha)}
  \left(k, l, -\frac{k_h^2}{m},
   \frac{\omega R}{c_p \bar \theta^{(0)} }\right)\, ,
\end{align}
and dispersion relation
\begin{align}
  \label{omega_mid-freq}
 \omega^2 = \bar N^2 \frac{ k^2 + l^2}{m^2 + (1-\beta)(k^2 + l^2)} \, ,
\end{align}
where the Brunt-Vaisala background frequency $\bar N$ and the
buoyancy amplitude $\tilde b^{(1+\alpha)}$ are defined in
  \eq{def_N_bar} and \eq{def_b}, respectively.  \revision{The complete
    derivation can be found in App.~B.}
\subsection{Single parcel model approximation and single monochromatic GW}
\label{sec_monochromatic_GW}
In the following we adopt a Lagrangian framework and consider the ice
physics of a single air parcel influenced by GW dynamics. Further, we
assume that the leading order vertical velocity in \eq{transf_dot_pi}
is solely due to a single GW and can be written as
\revision{
\begin{align}
  \label{w0_cos}
  w^{(0)}(\mathbf{x}(t_w), t_w) \approx w^{(0)}(\mathbf{x}(t_*), t_w) &= |\tilde w^{(0)}| \cos(\omega t_w + \phi)\, ,
\end{align}
with real amplitude $|\tilde w^{(0)}|$, phase
$\phi=\mathbf{x}(t_*) \cdot \mathbf{k} + \delta \phi$ and initial
position of the parcel $\mathbf{x}(t_*)$. In Sec.~\ref{superposition}
we generalize the approach for the case of superposition of many GW.}

\subsection{The different regimes in the ice dynamics}
The gravity wave dynamics changes the vertical velocity, pressure and
temperature fields in \eq{S} and hence leads to variations of $S$ and
consequently of $n$. Time series illustrating the qualitative behavior
of $S$ and $n$ under GW forcing are shown in Fig.~\ref{ref06m0}, see
the discussion in section~\ref{numerical_experiments} for details. A
typical situation observed is that $S$ fluctuates until it reaches (or
approaches sufficiently) the critical value $S_c$ at time $t_0$. At
$t_0$ the nucleation term in \eq{n} leads to an explosive production
of ice crystals. The increased number concentration $n$ implies a
reduction of $S$ below $S_c$ through the diffusional growth term in
\eq{S}. After this reduction $S$ continues to fluctuate due to the GW
forcing and might again approach $S_c$. Thus, in some cases we have
to consider ice nucleation in presence of pre-existing ice crystals,
which might be
suppressed under certain conditions.\\
Following the matched asymptotic approach of BS19, three different
regimes are considered here. First, the pre-nucleation regime with
$S < S_c$, where the dynamics takes place on the GW time scale. This
is followed by a nucleation regime, centered around time $t_0$ with
$S(t_0)=S_c$ and dynamics on the much faster nucleation time
scale. After the nucleation event the post-nucleation regime is
entered with $S < S_c$, characterized again by dynamics on the GW time
scale.

We observe that due to the assumption $\bar m = m_c$ the evolution of
$n$ and $S$ is decoupled from the one of $q$. Because of this we first
consider equations \eq{n} and \eq{S}. Once $n$ and $S$ are known, $q$
can be found from \eq{q}, \revision{see also the discussion in
  Sec.~\ref{derivation}\ref{summary_asy}}. The case where all three
equations \eq{n}, \eq{q} and \eq{S} are coupled is discussed in
Sec.~\ref{sec_var_mass}.

\subsection{Pre- and post-nucleation regime}
Because the dynamics in the pre- and post-nucleation regime takes
place at the same characteristic time scale, we treat them
simultaneously here. In both regimes $S$ is below the critical
value, $S < S_c$, and $S-S_c=\Order{1}$ even in the limit
$\eps \to 0$.  This implies that the nucleation term in \eq{n} is
transcendentally small
\begin{align}
\frac{J^*}{\eps} \exp \left(\frac{B^*}{\eps^2} (S -S_c) \right)
\to 0 \text{ for } \eps \to 0\, .
\end{align}
Next, we substitute in \eq{n}, \eq{S} the expansion \eq{expansion_ice}
for $n$, $S$ and collect the leading order terms.  Evaluating the
resulting equations at $z_{00}$ gives
\begin{align}
  \label{n_pre_post_dot}
  \dd{n^{(0)}}{t_w} &= 0\, ,\\
  \label{S_pre_post_dot}
 \dd{S^{(0)}}{t_w} &= \revision{- D^* (S^{(0)} -1 ) n^{(0)}+  S^{(0)} F^*(t_w)\, ,}
\end{align}
where \eq{bar_pT_z00}, \eq{transf_dot_pi}, \eq{w0_cos}, \eq{pi_zero}
are used and the GW forcing term is defined as
\begin{align}
  \label{def_a}
  \revision{  F^*(t_w) =  \frac{R L^* |\tilde w^{(0)}| }{c_p} \cos(\omega t_w + \phi) }
\end{align}
From \eq{n_pre_post_dot} one obtains that the number concentration
does not change with time 
\begin{align}
  \label{n_pre_post}
  n^{(0)} = \begin{cases}
             N_{pre}\quad  &\text{ in the pre-nucleation regime}\\ 
             N_{post}\quad &\text{ in the post-nucleation regime}              
            \end{cases}\, .            
\end{align}
Whereas the constant $N_{pre}$ is typically given by the initial
condition, the number concentration after the nucleation, $N_{post}$,
is at this stage unknown.  By integrating \eq{S_pre_post_dot} from the
initial time $t_*$ up to $t_w$, one obtains an integral representation
for $S^{(0)}$
\begin{align}
  \label{S_pre_post}
  S^{(0)}(t_w) &= S_*S_h(t_w,t_*)
                 + \int \limits_{t_*}^{t_w} dt' D^* n^{(0)} S_h(t_w,t') \, ,
\end{align}
where the propagator $S_h(t_w,t_*)$ is defined as
\revision{
\begin{align}
\label{S_h}  
  S_h(t_w,t_*) &= \exp\left\{- D^* n^{(0)}(t_w - t_*) + \int\limits_{t_*}^{t_w} dt  F^*(t) \right\}
\end{align}
}
and the constant $S_*=S(t_*)$ is given by the initial condition for the saturation ratio.
\subsection{Nucleation regime}
The nucleation regime is around the (unknown) time $t_0$, when the pre-nucleation
$S^{(0)}$ reaches the critical value $S_c$
\begin{align}
  \label{s0_nuc}
  S^{(0)}(t_0) = S_c\, ,
\end{align}
and it is characterized by the condition $S-S_c=\Order{\eps^2}>0$. The
$\eps^{-2}$ scaling in the exponent of the nucleation term in \eq{n}
indicates that the dynamics during nucleation evolves on a fast time
scale. This motivates to introduce a rescaled time coordinate
\begin{align}
 \tau &= \frac{t_w - t_0}{\eps^2}\, .
\end{align}
Equations \eq{n} and \eq{S} are expressed in terms of $\tau$ giving
\begin{align}
  \label{n_tau_1ode}
 \dd{n}{\tau} &= \eps J^* \exp \left(\frac{B^*}{\eps^2} (S -S_c) \right) + 
\eps^{4}\frac{S^*_n}{\rho} \pp{}{z_w} \rho  n  \\
\label{S_tau_1ode}
 \dd{S}{\tau} &= \eps^2\left[ - D^* (S-1) T n 
               -\frac{S}{\pi}\left(\eps^{2+\alpha}\dd{\pi'}{t_w} + \eps w\dd{\bar \pi}{z_s} \right)
               \left(\frac{L^*}{\eps T} -
               \frac{c_p}{R}\right)
               - \eps^{4}\frac{p}{p_{si}}J^* \exp \left\{\frac{B^*}{\eps^2} (S -S_c) \right\}\right]\, .
\end{align}
The last two equations can be written in compact form as
%
\begin{align}
  \label{n_tau_1ode_ord}
 \dd{n}{\tau} &= \eps J^* \exp \left(\frac{B^*}{\eps^2} (S -S_c) \right) + 
\Order{\eps^4}  \\
\label{S_tau_1ode_ord}
 \dd{S}{\tau} &= \eps^2\left[ - D^* (S-1) T n 
 - \frac{L^* S w}{\bar \pi} \dd{\bar \pi}{z_s}
 \right] +\Order{\eps^3}\, ,
\end{align}
where the nucleation term in \eq{S_tau_1ode} was eliminated using
\eq{n_tau_1ode}. From the leading order of \eq{S_tau_1ode_ord} one
obtains that the saturation ratio $S^{(0)}$ does not change during
nucleation
\begin{align}
  \label{s_eq_sc}
  \dd{S^{(0)}}{\tau} = 0 \rightarrow S^{(0)}(\tau)=S_c\, , 
\end{align}
consistent with \eq{s0_nuc}.
\revision{As shown in Appendix~C} the leading order number concentration,
$n^{(0)}$, satisfies
\begin{align}
  \label{def_m}
  \dd{n^{(0)}}{\tau} + \delta (n^{(0)})^2 - \gamma n^{(0)} = \mu \, ,
\end{align}
where the following constants are introduced 
\begin{align}
  \label{delta}
  \delta &= \frac{1}{2}B^*D^*(S_c - 1) \\
  \label{gamma}
\gamma &= \revision{F^*(t_0) B^* S_c} \, ,
\end{align}
and $\mu$ is some constant of integration.
Integrating \eq{def_m} from $0$ up
to some time $\tau$, one obtains for the number concentration in the
nucleation regime
\begin{align}
  \label{n_tau}
  n^{(0)}(\tau) = \frac{n_s + n_e C e^{\sigma \tau}}{1 + C e^{\sigma \tau}}\, ,
\end{align}
with the constants
\begin{align}
  \label{sigma} 
  \sigma &= \sqrt{\gamma^2 + 4 \delta \mu} \\
  n_s &= \frac{\gamma -\sigma}{2 \delta}\\
  \label{def_ne}
  n_e &= \frac{\sigma + \gamma}{2 \delta}\\
  \label{def_C}
  C &= \frac{n_0 - n_s}{n_e - n_0}
\end{align}
and another constant of integration $n_0 = n^{(0)}(0)$.
\subsection{Matching}
\label{matching}
Next, we find the constants of integration, entering the solution in
the nucleation and the post-nucleation regime by matching the
different solutions in the pre-nucleation, nucleation and
post-nucleation zone.  First, we consider the limits
$\tau \to \pm \infty$ of the nucleation solution \eq{n_tau}
\begin{align}
  \label{n_tau_minfty}
  n^{(0)}_{nuc}(-\infty) &=  n_s\\
  \label{n_tau_infty}
  n^{(0)}_{nuc}( \infty) &=  n_e\\
  \label{n_tau_dot}
  \dd{n^{(0)}_{nuc}}{\tau}(\pm \infty) &= 0\, .
\end{align}
The nucleation solution for the number concentration, $n^{(0)}_{nuc}(\tau)$,
should match for $\tau \to -\infty$ the one from the pre-nucleation
regime, $n^{(0)}_{pre}(t)$, for $t \to t_0$. Equating
\eq{n_tau_minfty} and \eq{n_pre_post} give
\begin{align}
  \label{ns}
  n_s &= N_{pre}\, .
\end{align}
By considering \eq{def_m} for $\tau \to -\infty$
one obtains with the help of
\eq{n_tau_minfty}, \eq{n_tau_dot} and \eq{ns}
\begin{align}
  \label{value_m}
  \mu &=  \delta N_{pre}^2 - \gamma N_{pre}\, . 
\end{align}
Matching the saturation ratio in the nucleation regime,
$S^{(0)}_{nuc}$ from \eq{s_eq_sc}, to the one in the pre-nucleation
regime, $S^{(0)}_{pre}$ from \eq{S_pre_post}, gives the
condition
\begin{align}
  \label{value_t0}
  S_* S_h(t_0,t_*)
  + \int \limits_{t_*}^{t_0} dt' D^* N_{pre} S_h(t_0,t')
  = S_c\, .
\end{align}
The last equation is an implicit equation for the time of the
nucleation
event $t_0$, where \revision{$S_h(t_0,t_*)=\exp\left\{- D^* N_{pre}(t_0 - t_*) + \int\limits_{t_*}^{t_0} dt  F^*(t) \right\} $} and $S_*$ is given by the initial condition.\\
Next, the nucleation solution for the number concentration, $n_{nuc}(\tau)$, should match for $\tau \to \infty$ the one from the post-nucleation regime,
$n^{(0)}_{post}(t)$, for $t \to t_0$. Equating \eq{n_tau_infty} and  \eq{n_pre_post} gives
\begin{align}
  n_e = N_{post}\, .
\end{align}
From \eq{def_ne}, with $\sigma$ and $\mu$ given by \eq{sigma} and \eq{value_m}, respectively, the post-nucleation value for the number concentration can be found
\revision{
\begin{align}
  \label{n_final}
  N_{post}&=
  \begin{cases}
   \frac{2 F^*(t_0) S_c }{D^* (S_c -1)} - N_{pre}
   \quad &\text{ if } N_{pre} < \frac{F^*(t_0) S_c }{D^* (S_c -1)} \\
   N_{pre} &\text{ else }
\end{cases}
    \, .
\end{align}
}
The two cases in the last solution result from the condition that the
expression under the root in \eq{sigma} is positive. Eq \eq{n_final}
implies that there are newly nucleated ice crystals only for
$N_{pre} < N^{c}_{pre}=\frac{F^*(t_0) S_c }{D^* (S_c -1)}$. In the
case of nucleation the larger $N_{pre}$ the smaller $N_{post}$ is,
however the average $(N_{pre}+N_{post})/2$ does not depend on the
initial $n$ and is always the same (for fixed
$F^*(t_0)$). Interestingly, the threshold $N^{c}_{pre}$ has important
implication for the pre-nucleation dynamics of the saturation
ratio. For $N_{pre} > N^{c}_{pre}$ most likely the necessary condition
$S=S_c$ for the existence of nucleation will not be met. To see this
consider that $S$ is increasing just before nucleation, so one has
$\dot S > 0$ at time $t_0$ when $S=S_c$. From \eq{S_pre_post_dot} this
leads again to the condition $N_{pre}<N^{c}_{pre}$ for a nucleation,
if $\cos(\omega t_0 + \phi)>0$ is assumed. Eq. \eq{n_final} defines
further two interesting limits \revision{
\begin{align}
  \label{n_max_post}
  N^{max}_{post} &= \frac{2 F^*_{max} S_c}{D^*(S_c-1)} - N_{pre}  \\
  \label{n_max_pre}
  N^{max}_{pre} &  = \frac{F^*_{max} S_c}{D^*(S_c-1)} \, ,
\end{align}
where $F^*_{max} = \max\{F^*(t): t \in [0, 2\pi T_w]
\}$.} $N^{max}_{post}$ is the maximum possible post-nucleation number
concentration due to a GW with a given amplitude. $N^{max}_{pre}$ is
the maximum possible pre-nucleation number concentration which might
allow for a nucleation \citep[see also the discussion on pre-existing
ice in][]{gierens2003}.

\subsection{Summary of the reduced model}
The asymptotic analysis presented here allows to identify a reduced
model for the dominant interactions of the ice physics with the GW
dynamics. Here we summarize it since the model will be used for the
evaluation in the next Section. It contains only the dominant terms
from the full ice physics model \eq{n}-\eq{S} evaluated at level
$z_{00}$. In dimensional form it reads
\revision{
\begin{align}
\label{n_ref}
  \dd{n}{t} &= J \exp \left( B (S -S_c) \right)\, , \\
\label{S_ref}
  \dd{S}{t} & = - D (S-1) n + S F(t)\, , \\
  \label{q_ref}
  \dd{q}{t} & =  D (S-1) n \, ,
\end{align}}
with $F$ defined in  \eq{A_def}.

\subsection{Summary of the asymptotic solution}
\label{summary_asy}
\revision{In Appendix~D we construct} the composite asymptotic solutions
\eq{n_comps_ndim} and \eq{S_comps_ndim} valid in all three different
regimes. Using the replacements
\begin{align}
  T_w (t_w, t_0) &\to (t,  t_0)\\
  n_c (n, n_s, n_e) &\to ( n,  n_s,  n_e)
\end{align}
the equations are re-dimensionalized giving
\begin{align}
  \label{n_asy}
  n(t) &= \frac{n_s + n_e e^{ \zeta (t - t_0)}}
              {1 + e^{\zeta (t - t_0)}}\, , &\\
  \label{S_asy}
  S(t) &=
  \begin{cases}
    S_*S_h(t,t_*) + \int \limits_{t_*}^{t} dt' D n_s
    S_h(t,t') \qquad &\text{ for }  t \le t_0 \\
    S_c S_h(t, t_0) + \int \limits_{t_0}^{ t} dt' D n_e
    S_h( t, t') \qquad &\text{ for }  t >  t_0
  \end{cases}  
\end{align}
with the definitions
\revision{
\begin{align}
  \zeta &=  B S_c F(t_0) - B D(S_c -1)n_s\\
  \Omega &= \frac{\omega}{T_w}\, , 
  D = \frac{D^*}{T_w n_c}\\
  \label{A_def}
   F(t_0) &=  \frac{g L_i \hat w}{c_p R_v T_{00}^2} \cos(\Omega t_0 + \phi)
\end{align}
}
\revision{ and the dimensional GW vertical velocity amplitude
  $\hat w$.  Note further that the final post-nucleation number
  concentration $n_e$ is given in dimensional form by (\ref{n_final})
  with all asterisks omitted and the nucleation time, $t_0$, is found
  from the condition $S(t_0)=S_c$.}

\revision{In order to account for conservation of total water we
  supplement the asymptotic system with the equation for the ice
  mixing ratio \eq{q_ref}.  By integrating the latter equation with
  $S$ from \eq{S_asy} and constant $n=n_s$, one has an integral
  representation for $q$ in the pre-nucleation regime
\begin{align}
 q(t) =  \int \limits_{t_*}^t D n_s (S(t')-1) dt' + q(t_*)\, .
\end{align}
For initial $q(t_*)>0$, $q$ vanishing at some later time,
$t_{\text{evap}}$, implies that all initial crystals evaporated. The
solutions \eq{n_asy}, \eq{S_asy} can still be used after
$t_{\text{evap}}$, if the pre-nucleation value $n_s$ is set to
zero at $t_*=t_{\text{evap}}$. The post-nucleation solution can be
treated in a similar way to account for evaporation events.}

\section{Numerical experiments and discussion of the asymptotic
  solution}
\label{numerical_experiments}

\begin{figure}[h!]
  \centering
  \includegraphics[width=0.45\textwidth]{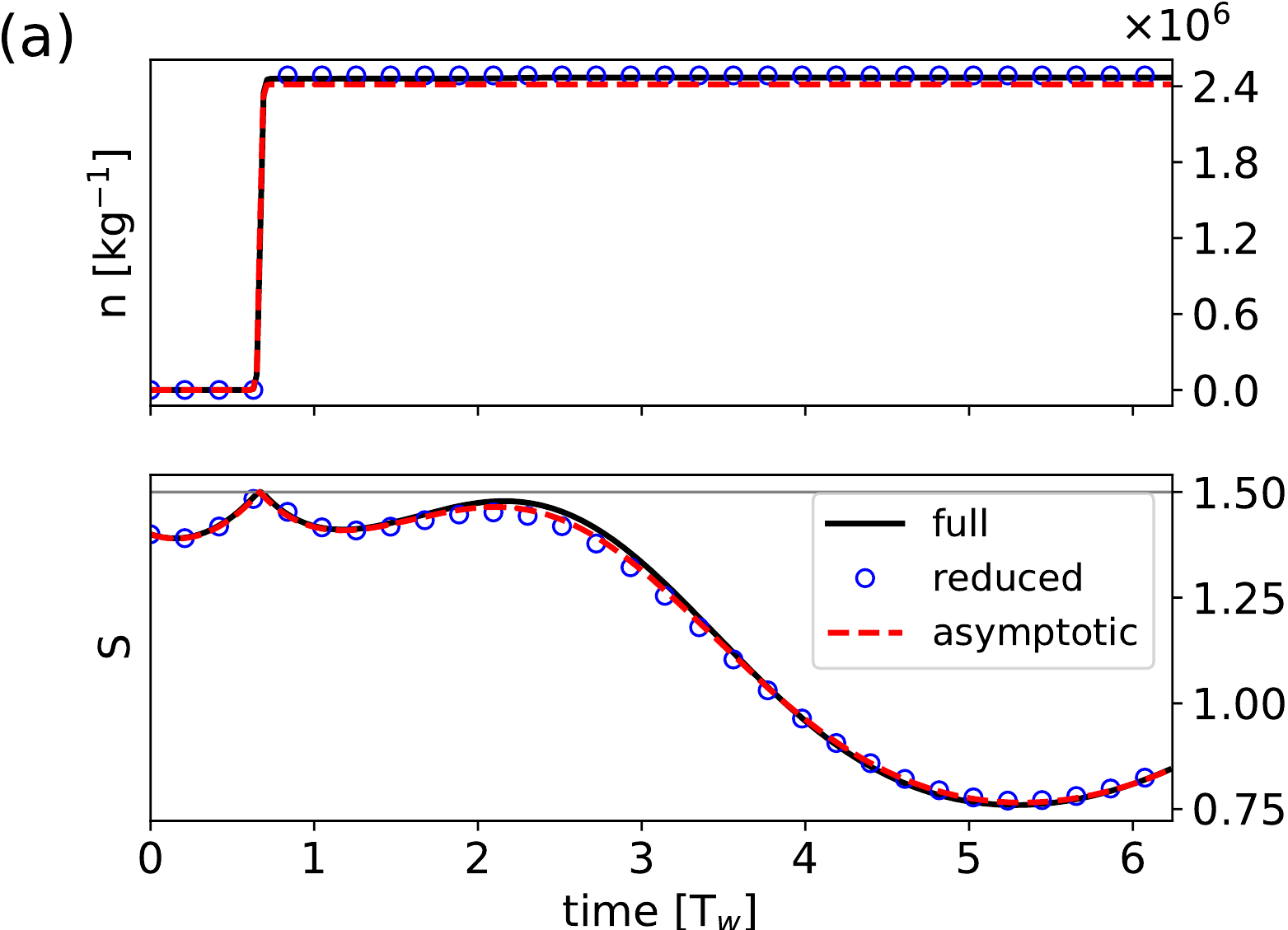}
  \includegraphics[width=0.45\textwidth]{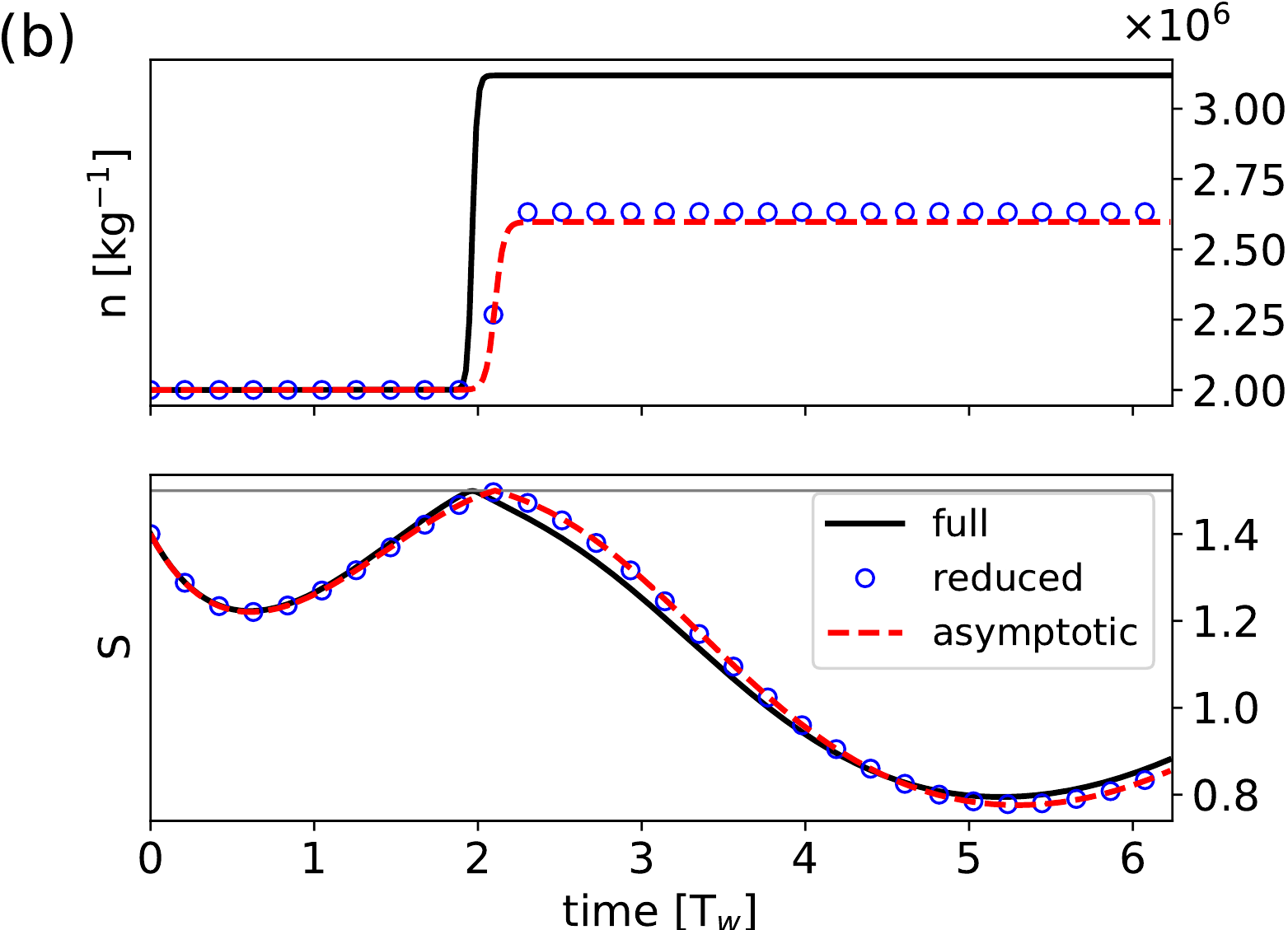}
  \caption{\revision{Time evolution of number concentration $n$ and
      saturation ratio $S$ for two different inital conditions:
      $n(0)=0$ (a) and $n(0)=2\times 10^{6}$ kg$^{-1}$ (b). $T_w=500$ s
      and initial phase of the wave $\phi=-\frac{11\pi}{20}$}}
  \label{ref06m0}
\end{figure}

\begin{figure}[h!]
      \centering
      \includegraphics[width=0.49\textwidth]{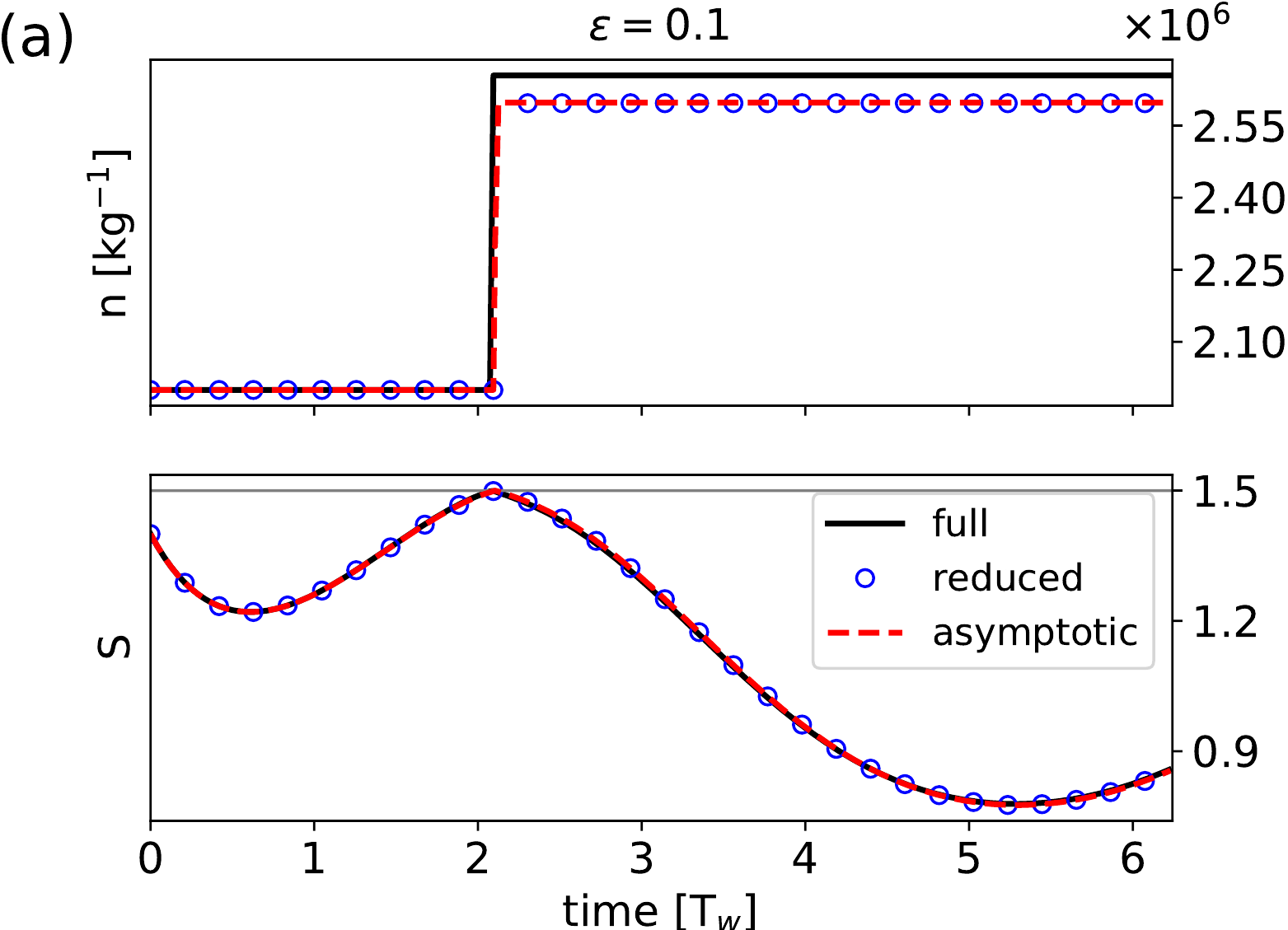}
      \includegraphics[width=0.49\textwidth]{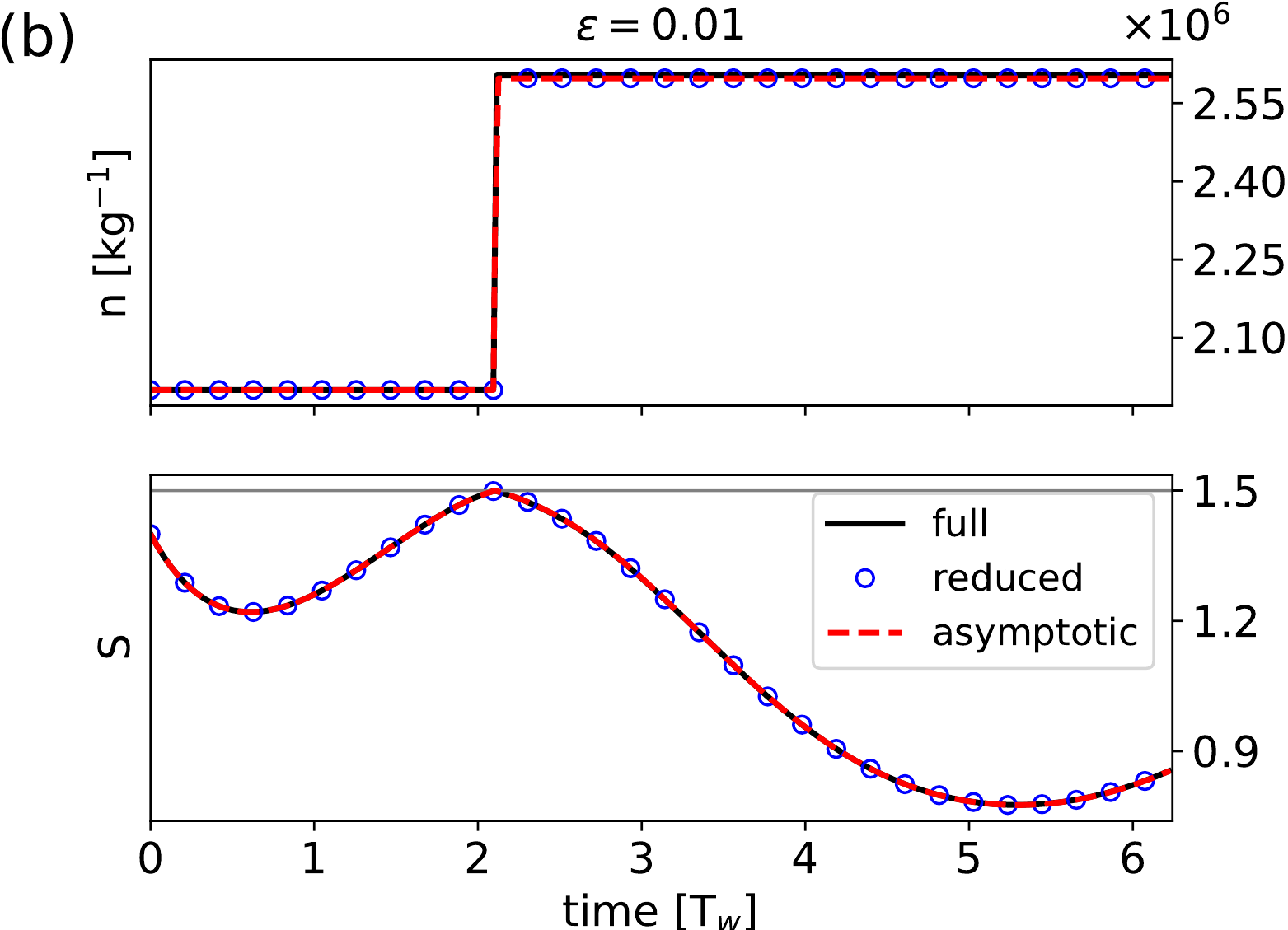}
      \caption{As in Fig.~\ref{ref06m0}b but for $\eps=10^{-1}$ (a) and $\eps=10^{-2}$ (b); in Fig.~\ref{ref06m0}b $\eps=1$. }
\label{fig_asym_limit}
\end{figure}

\revision{In this Section the asymptotic model is validated against
  the full ice microphysics model and the reduced model for realistic
  parameter values taken from BS19 and summarized in
  tab.~\ref{tab:ref2}. The full model solves \eq{dot_n_non}-\eq{q_v}
  omitting only sedimentation, the details of it can be found in
  App.~E. All models are forced with single monochromatic GW from
  Sec.~\ref{derivation}\ref{sec_monochromatic_GW}. The GW vertical
  velocity amplitude is set to $1$ ms$^{-1}$ corresponding to
  $0.7 W_c$, where $W_c$ is the critical vertical velocity amplitude
  for breaking due to static instability. The GW frequency is
  $\Omega={T_w}^{-1} =2\times 10^{-3}$ s$^{-1}$ and the initial
  saturation ratio is set to $S(0)=1.4$

  The results for two different initial number concentrations are
  summarized in Fig.~\ref{ref06m0}. Figure~\ref{ref06m0}a shows a
  situation where the asymptotic solution reproduces with a
  high-accuracy the time evolution of the ice crystal number
  concentration and saturation ratio. Fig.~\ref{ref06m0}b depicts a
  case where the nucleation time $t_0$ from the full model is slightly
  missed by the asymptotic and the reduced models. Although the overall
  evolution of $S$ is reproduced well, the nucleated ice crystal
  number is underestimated by roughly 15 \%. From the asymptotic
  theory, we expect that the discrepancy will vanish in the limit
  $\eps \to 0$.  This asymptotic limit is verified numerically by
  considering smaller values of $\eps$ in the full and in the reduced
  model, this corresponds to increasing the time scale separation
  between the different processes in the models. The results are
  summarized in Fig.~\ref{fig_asym_limit} for $\eps=10^{-1}$ and
  $\eps=10^{-2}$, note that in Fig.~\ref{ref06m0} $\eps=1$ implying no
  increased time scale separation. Fig.~\ref{fig_asym_limit} shows
  that the models converge quickly to the asymptotic limit already for
  moderately small values of $\eps$.}

\begin{figure}[h!]
  \centering
   \includegraphics[width=0.5\textwidth] {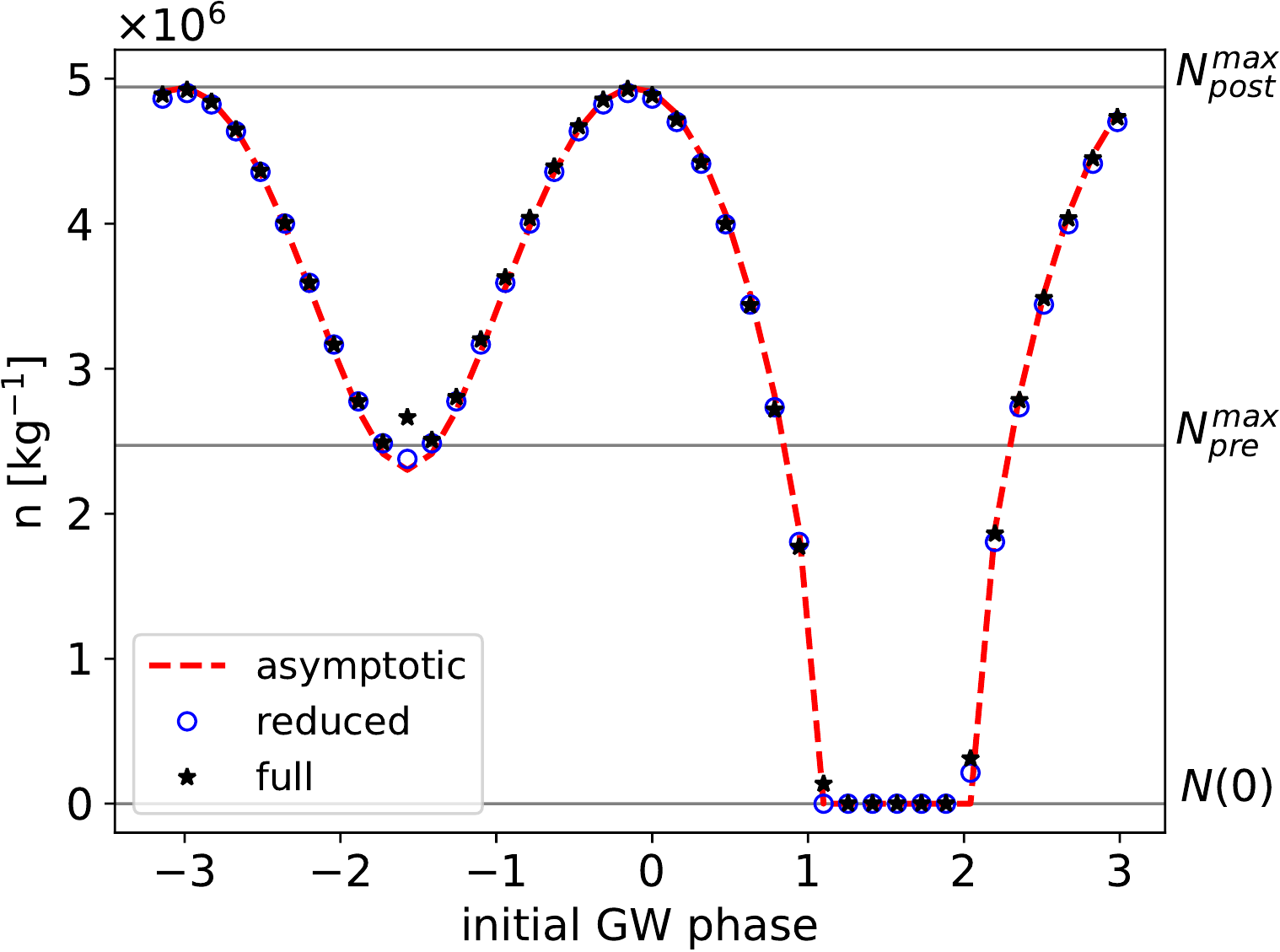}
  \caption{\revision{Nucleated number concentration, $n$, as a function of
      the initial GW phase. Grey horizontal lines denote the initial
      condition $n(0)$ and the asymptotic estimates $N^{max}_{pre}$
      and $N^{max}_{post}$ from equations \eq{n_max_post} and
      \eq{n_max_pre}, respectively.}}
  \label{ini_time}
\end{figure}

\revision{Since the phase of the GW is typically unknown in coarse
  models, we study the sensitivity of the results with respect to this
  parameter. For that purpose we vary the GW phase at the initial time
  $t=0$ and determine the nucleated ice crystals within one wave
  period for an initial condition $n(0)=0$. From Fig.~\ref{ini_time}
  it is visible that the asymptotic solution captures for all GW
  phases the number of nucleated ice crystals in the full model. In
  addition, the values of $n$ are limited from above by the asymptotic
  estimate $N^{max}_{post}$.}
%

\begin{figure}[h!]
  \centering
   \includegraphics[width=0.5\textwidth] {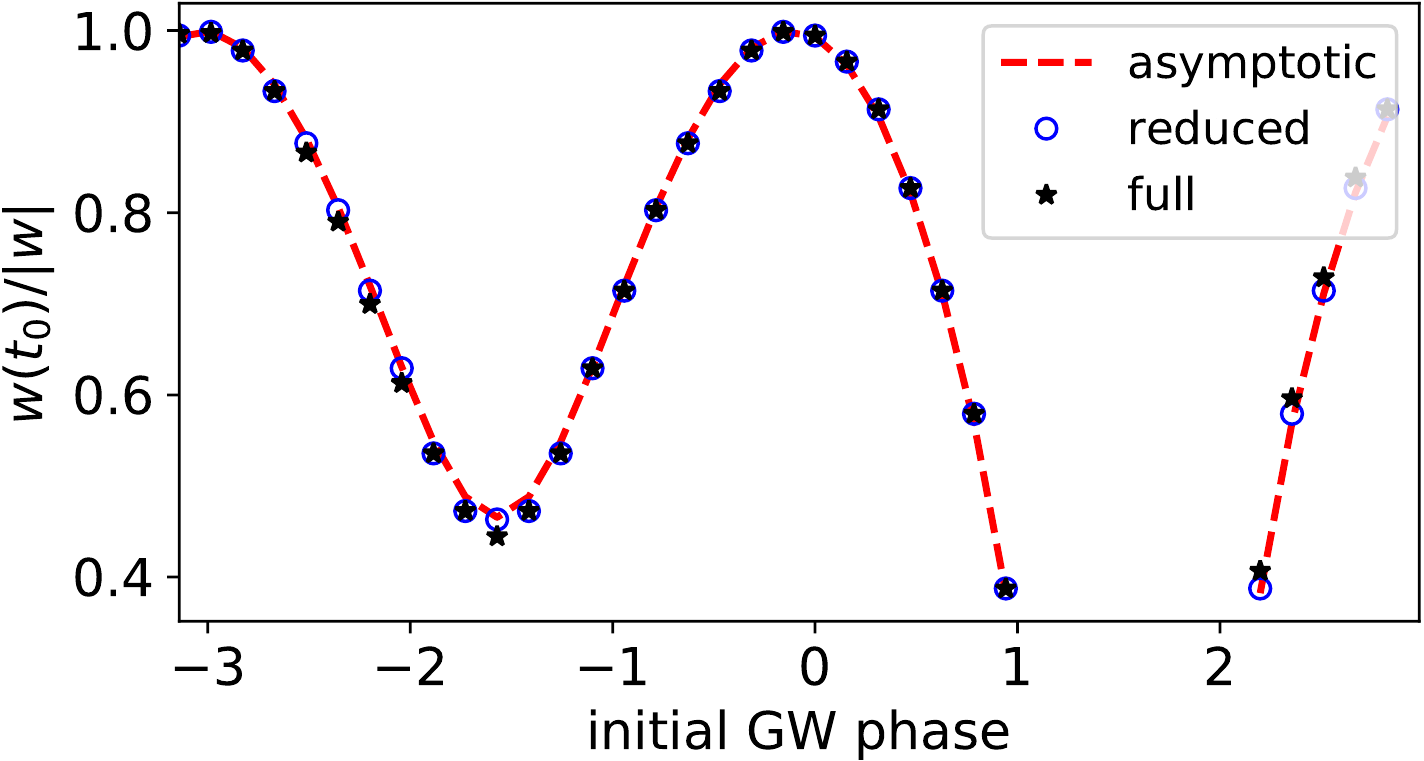}
  \caption{\revision{Normalized GW vertical velocity at $t_0$ as a
      function of the initial GW phase for the simulations from
      Fig.~\ref{ini_time}.}}
  \label{w_ini_time}
\end{figure}

\revision{In Fig.~\ref{w_ini_time} the normalized vertical velocity at
  the nucleation time is displayed. It suggests that nucleation takes
  place at sufficiently high updrafts but not necessary at the
  maximal.}

\begin{figure}[h!]
  \centering
  \includegraphics[width=0.49\textwidth]{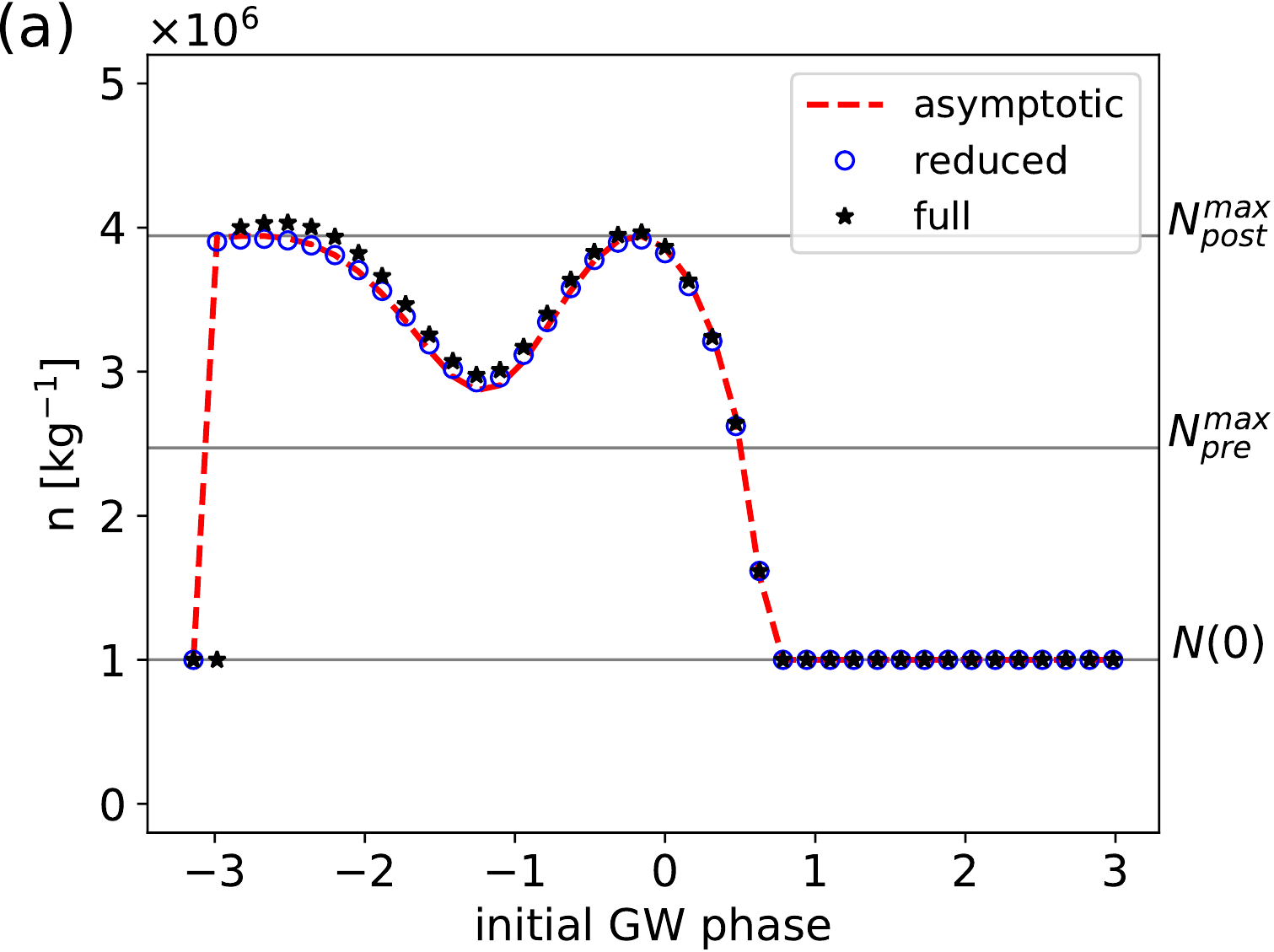}
   \includegraphics[width=0.49\textwidth]{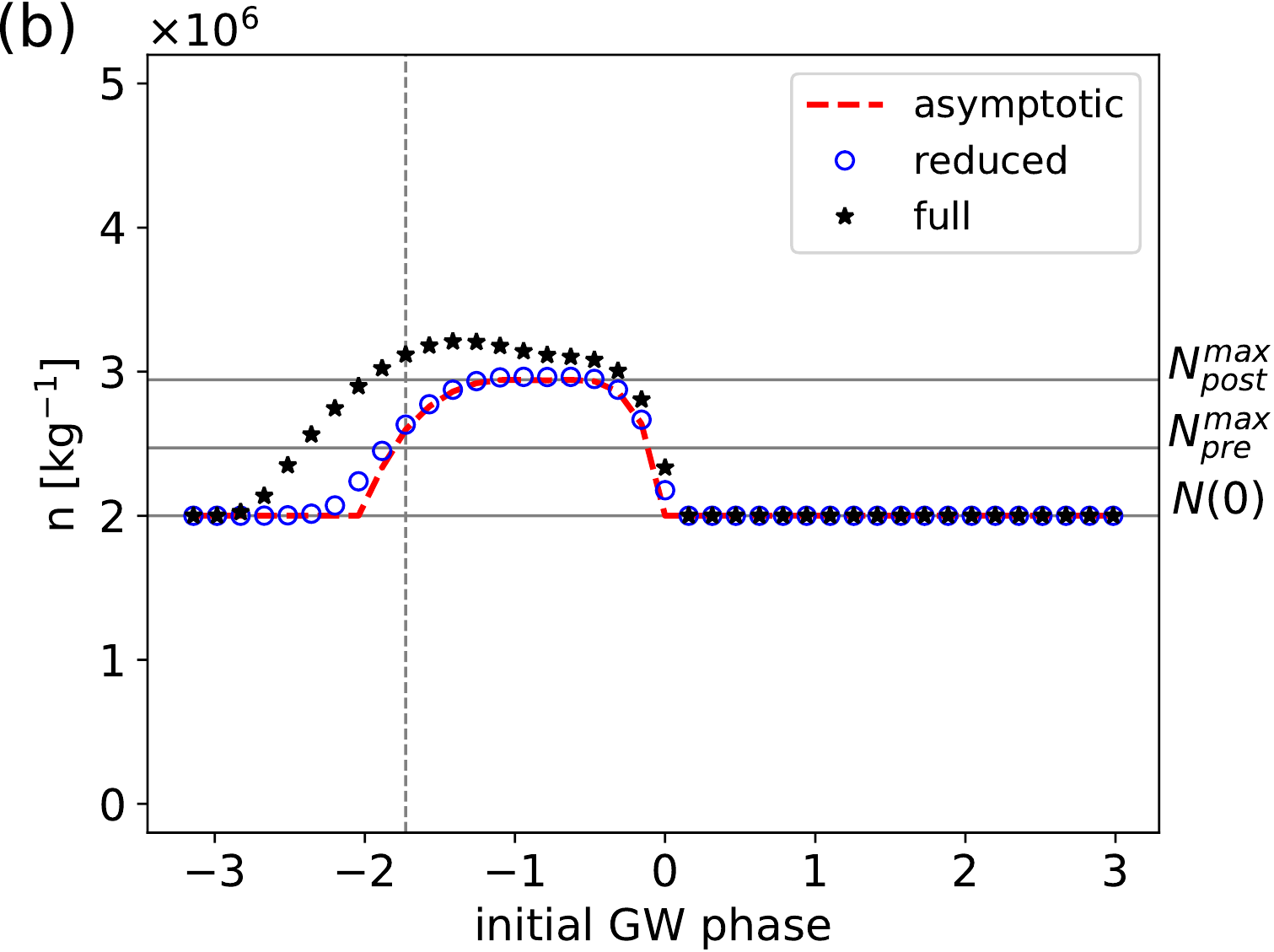}
    \caption{\revision{As in Fig.~\ref{ini_time} but for initial
        number concentration $n(0)=10^{6}$ kg$^{-1}$ (a) and
        $n(0)=2\times 10^{6}$ kg$^{-1}$ (b).  In Fig. (b) the initial
        $n$ is the same as the one used in Fig.~\ref{ref06m0}b and
        \ref{fig_asym_limit}; the dashed gray vertical line marks the
        initial phase used for the simulation in Fig.~\ref{ref06m0}b
        and \ref{fig_asym_limit}.}}
  \label{ini_time_ic}
\end{figure}

\revision{Fig.~\ref{ini_time_ic} summarizes the dependence of the
  nucleated ice crystals on the initial number concentration. The
  asymptotic solution reproduces nearly exactly $n$ from the reduced
  model. Both models are very close to the full model for $n(0)$ well
  below $N^{max}_{pre}$, however, for $n(0)$ approaching
  $N^{max}_{pre}$ they underestimate $n$ for some GW phases. The
  detailed evolution of the solutions for one such particular case
  (dashed line in Fig.~\ref{ini_time_ic}b) was shown in
  Fig.~\ref{ref06m0}b. As shown in Fig.~\ref{fig_asym_limit} the
  discrepancy in the models results from the finite time scale
  separation between processes and diminishes for $\eps \to
  0$. Further, one observes in Fig.~\ref{ini_time_ic} that the
  asymptotic solution is able to capture regimes without nucleation
  events, too.}

\section{The effect of variable ice crystal mean mass \revision{in the
    deposition}}
\label{sec_var_mass}
Performing realistic air parcel simulations with a box model and a
bulk microphysics scheme, BS19 demonstrated that for a wide variety of
environmental conditions the \revision{constant mean mass assumption
  in the deposition term} is a reasonable approximation during
nucleation. However, right before a nucleation event where the
saturation ratio is above one, the mean mass of the ice crystals will
grow leading to an increased deposition term, see the $\bar m$
dependence in \eq{dep}. The increased deposition will influence the
saturation ratio, which on the other hand will affect the number of
nucleated ice crystals. Since the mean mass of the ice crystals can be
diagnosed from the relation $m=q/n$ for $n>0$, such effects can be
incorporated in the present model if the substitution
$\bar m=m(t)=q/n$ is introduced in the diffusional growth term. By
considering again only the dominant terms in the prognostic equations,
this results in the following system of reduced equations \revision{with
variable mean mass in the deposition}
\begin{align}
\label{n_m}
  \dd{n}{t} &= \frac{J^*}{\eps} \exp \left( \frac{B^*}{\eps^2} (S -S_c) \right) \\
\label{S_m}
 \dd{S}{t} &=\revision{ - D^* \left(\frac{q}{n} \right)^{\frac{1}{3}} (S-1) n + S F^*(t)}\, ,\\
\label{q_m}
 \dd{q}{t} &=  \frac{D^*}{\eps} \left(\frac{q}{n} \right)^{\frac{1}{3}} (S-1) n\, .
\end{align}
%

\begin{figure}[h!]
      \centering
      \includegraphics[width=0.5\textwidth]{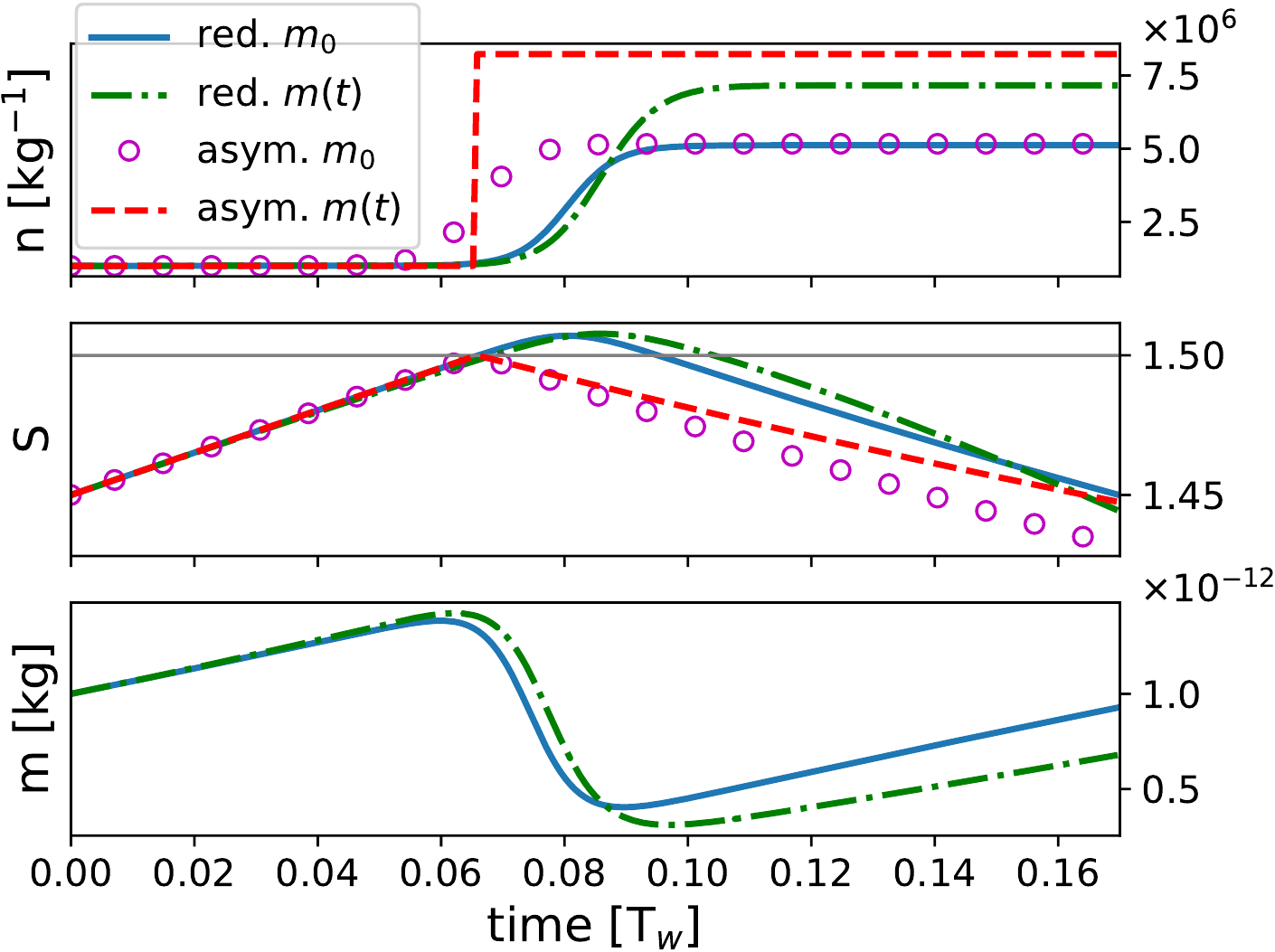}
      \caption{\revision{Time evolution of number concentration $n$,
          saturation ratio $S$ and mean ice mass $m$ computed for
          different models: the reduced model with constant mass
          \eq{n_ref}-\eq{q_ref}, the reduced model with variable mass
          \eq{n_m}-\eq{q_m}, the asymptotic solution for constant mass
          Sec.~\ref{derivation}\ref{summary_asy} and the asymptotic
          solution with variable mass correction
          \eq{alg_eq_n_post}. Note that for the latter model only the
          pre- and post-nucleation values are plotted, resulting in
          jump at $t_0$. The mean mass in the reduced models is
          diagnosed using $m=q/n$. In the figure's legend $m_0$ and
          $m(t)$ denote models with constant and variable mass,
          respectively.}}
\label{mtsngl}
\end{figure}

In Fig.~\ref{mtsngl} we show simulations of the reduced model with
constant and with variable mean mass, in both cases the ice crystal
mass is diagnosed using $m=q/n$. \revision{The vertical velocity
  amplitude is set to $1.25$ ms$^{-1}$ corresponding to $0.9 W_c$. The
  initial conditions for all models are set to $S(0)=1.45$,
  $n(0)= 10^{6}$ kg$^{-1}$ and $\phi=0$.}

In Fig.~\ref{mtsngl} an increase of the ice crystal mass is observed
before the nucleation event. Taking this into account with the model
\eq{n_m}-\eq{q_m} results in larger number of nucleated ice particles,
as compared to the constant mass model. \revision{The asymptotic
  solution for the constant mass case (denoted with ``asym. $m_0$ ``
  in the Figure) reproduces the behavior of the constant mass reduced
  model and underestimates $n$, as well.} In the following, we extend
the asymptotic approach to allow for variable mean mass effects.

\subsection{Pre-nucleation regime}
\revision{The fully coupled system \eq{n_m}-\eq{q_m} involves an
  additional fast time scale in the $q$-equation as compared to the
  constant mean mass case discussed in Sec.~\ref{derivation}. The new
  time scale will induce in general nontrivial dynamics on longer time
  scales, however, the corresponding rigorous asymptotic analysis is
  out of the scope of this paper. As we will show here, the asymptotic
  pre-nucleation solution for the constant mass case can still be used
  for variety of configurations if appropriate corrections are
  introduced.}

First, we observe in Fig.~\ref{mtsngl} that before nucleation the
solution for $S$ does not change much if the variable mean mass effect
is taken into account. \revision{In such situations, one can use} the
asymptotic solution \eq{S_pre_post} for $S$ to compute the time,
$t_0$, of the nucleation event from \eq{value_t0}. For
$S-S_c = \Order{1}<0$ the right-hand-side of \eq{n_m} vanishes at
leading order, implying the constant solution $n(t)=N_{pre}$. Using
the latter result and integrating \eq{q_m} from $t_*$ up to $t_0$, we
obtain at the end of the pre-nucleation regime
\begin{align}
  \label{q_t0}
  q(t_0)^{\frac{2}{3}} = \frac{3D^*}{2 \eps} N_{pre}^{\frac{2}{3}} \int \limits_{t_*}^{t_0} dt' (S(t')-1)+q(t_*)^{\frac{2}{3}}\, ,
\end{align}
with $S(t)$ from \eq{S_pre_post}.
%
%

\subsection{Nucleation regime}
Next, we consider the nucleation regime. On the fast nucleation time
$\tau$ we have the following system of equations
\begin{align}
\label{n_m_tau}
  \dd{n}{\tau} &= \eps J^* \exp \left( \frac{B^*}{\eps^2} (S -S_c) \right) \\
\label{S_m_tau}
  \dd{S}{\tau} &=
  \revision{               
    - \eps^2 D^* \left(\frac{q}{n} \right)^{\frac{1}{3}} (S-1) n
                 + \eps^2 S F^*(t_0 + \eps^2 \tau)\, ,}\\
\label{q_m_tau}
 \dd{q}{\tau} &=  \eps D^* \left(\frac{q}{n} \right)^{\frac{1}{3}} (S-1) n
\end{align}
From \eq{q_m_tau} we see that the leading order ice mixing ratio is constant
during nucleation
\begin{align}
q^{(0)}(\tau) = q^{(0)}_{pre}(t_0)\, ,
\end{align}
%
with the value of $q^{(0)}_{pre}(t_0)$ from \eq{q_t0}. Next,
by repeating the manipulations in equations \eq{n_tau2} to \eq{n_dd},
one obtains from \eq{n_m_tau}, \eq{S_m_tau} the following evolution
equation for the number concentration
\begin{align}
  \label{def_m_m}
  \dd{n^{(0)}}{\tau} + \tilde \delta (n^{(0)})^{5/3} - \gamma n^{(0)} = \tilde \mu \, ,
\end{align}
with the constant
\begin{align}
\label{delta_m}  
  \tilde \delta &= \frac{3}{5}\left(q^{(0)}\right)^{1/3}B^*D^*(S_c - 1)
\end{align}
and $\gamma$ defined in \eq{gamma} (see below \eq{alg_eq_n_post} for
the definition of $\tilde \mu$). Eq \eq{def_m_m} can be solved
numerically to find the number concentration during the nucleation
regime. The numerical integration of the ODE can be avoided if only
the final $n$ at the end of the nucleation is of interest. Proceeding
as in Sec.~\ref{derivation}\ref{matching} we have the following
matching conditions for the pre-nucleation, nucleation and
post-nucleation regime
 \begin{align}
   n^{(0)}_{nuc}(-\infty) &= N_{pre}\, ,\\
   n^{(0)}_{nuc}(\infty) &= N_{post}\, ,\\ 
   \dd{n^{(0)}_{nuc}}{\tau}(\pm \infty) &=
   \dd{n^{(0)}_{pre}}{t}(t_0) =  \dd{n^{(0)}_{post}}{t}(t_0)  = 0
 \end{align}
 Inserting the matching conditions in \eq{def_m_m} leads to an algebraic equation
 for $N_{post}$
 \begin{align}
   \label{alg_eq_n_post}
  \tilde \delta N_{post}^{5/3} - \gamma N_{post} = \tilde \mu\, ,
\end{align}
where $\tilde \mu = \tilde \delta N_{pre}^{5/3} - \gamma N_{pre}$.
Note that for $N_{pre}=0$, \eq{alg_eq_n_post} implies for the
dependence of $N_{post}$ on the GW vertical velocity:
$N_{post}\sim \hat w^{3/2}$, which is consistent with the scaling in
\cite{karcher_parameterization_2002}.

\subsection{Numerical results}

Equation \eq{alg_eq_n_post} is used to find an asymptotic
approximation of the nucleated number concentration, the comparison
with the numerical results is shown in Fig.~\ref{mtsngl}. The current
procedure allows to produce larger number of nucleated ice crystals as
compared with the constant mass model, the magnitude of $n$ is close
to the one of the variable mean mass model.

\begin{figure}[h!]
  \centering
  \includegraphics[width=0.5\textwidth]{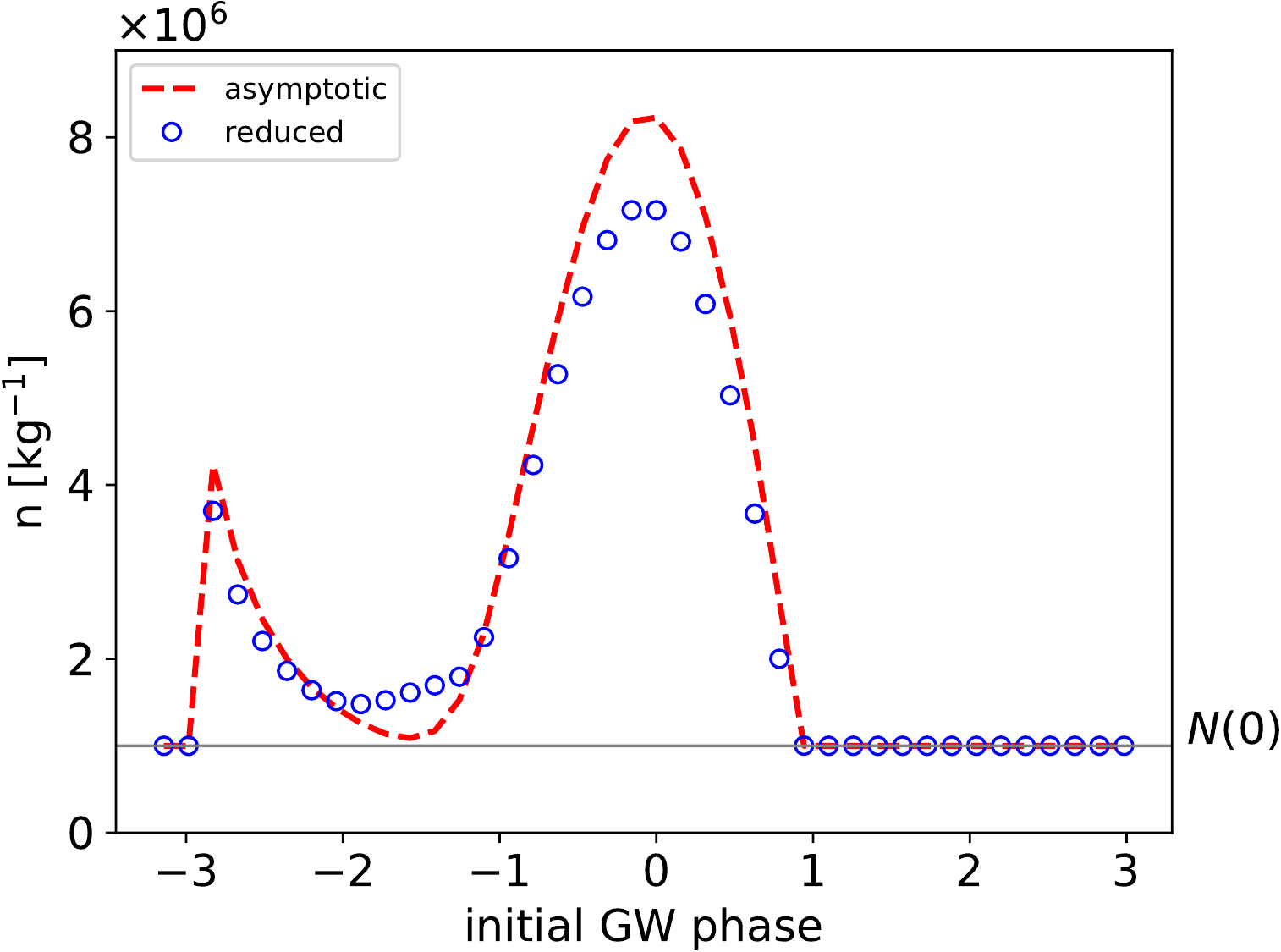}
  \caption{Number concentration $n$ as a function of the initial GW
    phase for the variable mean mass model \eq{n_m}-\eq{q_m}
    and for the asymptotic solution \eq{alg_eq_n_post} from
    Sec.~\ref{sec_var_mass}. Initial conditions as in
    Fig.~\ref{mtsngl}.}
  \label{mt}
\end{figure}

The performance of the current approach is systematically evaluated by
varying the initial GW phase. The corresponding results are summarized
in Fig.~\ref{mt}. The figure suggests that the proposed procedure
captures the number of ice crystals after nucleation for various
initial GW phases.

A note of caution should be added on the relevance of the variable
mean mass model presented in this section.  \revision{The use of the
  pre-nucleation $S$ from \eq{S_pre_post} in \eq{q_t0} might be not
  valid on longer time scales, see the first paragraph of Sec.~5a.}
Further, Fig.~\ref{mtsngl} suggests that there might be situations
with considerable growth of $m$ before nucleation takes
place. Obviously, for large $m$ the sedimentation term will provide a
sink for the mean mass: see \eq{q}. The correct incorporation of the
sedimentation effects will be subject of a future study.

\section{Ice physics forced by superposition of gravity waves}
\label{superposition}
\begin{figure}[h!]
  \centering
  \label{ensemble}
  \includegraphics[width=0.49\textwidth]{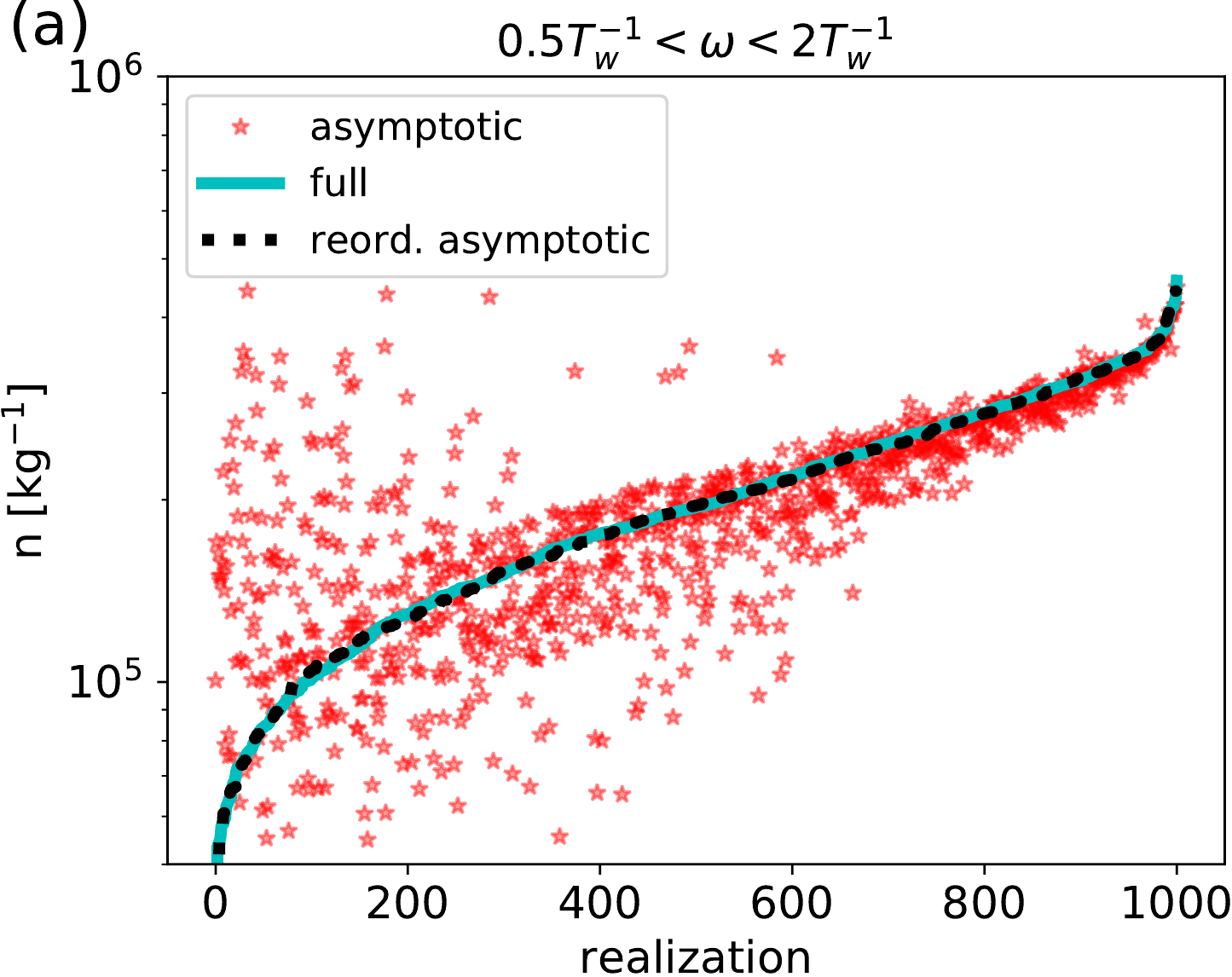} 
  \includegraphics[width=0.49\textwidth]{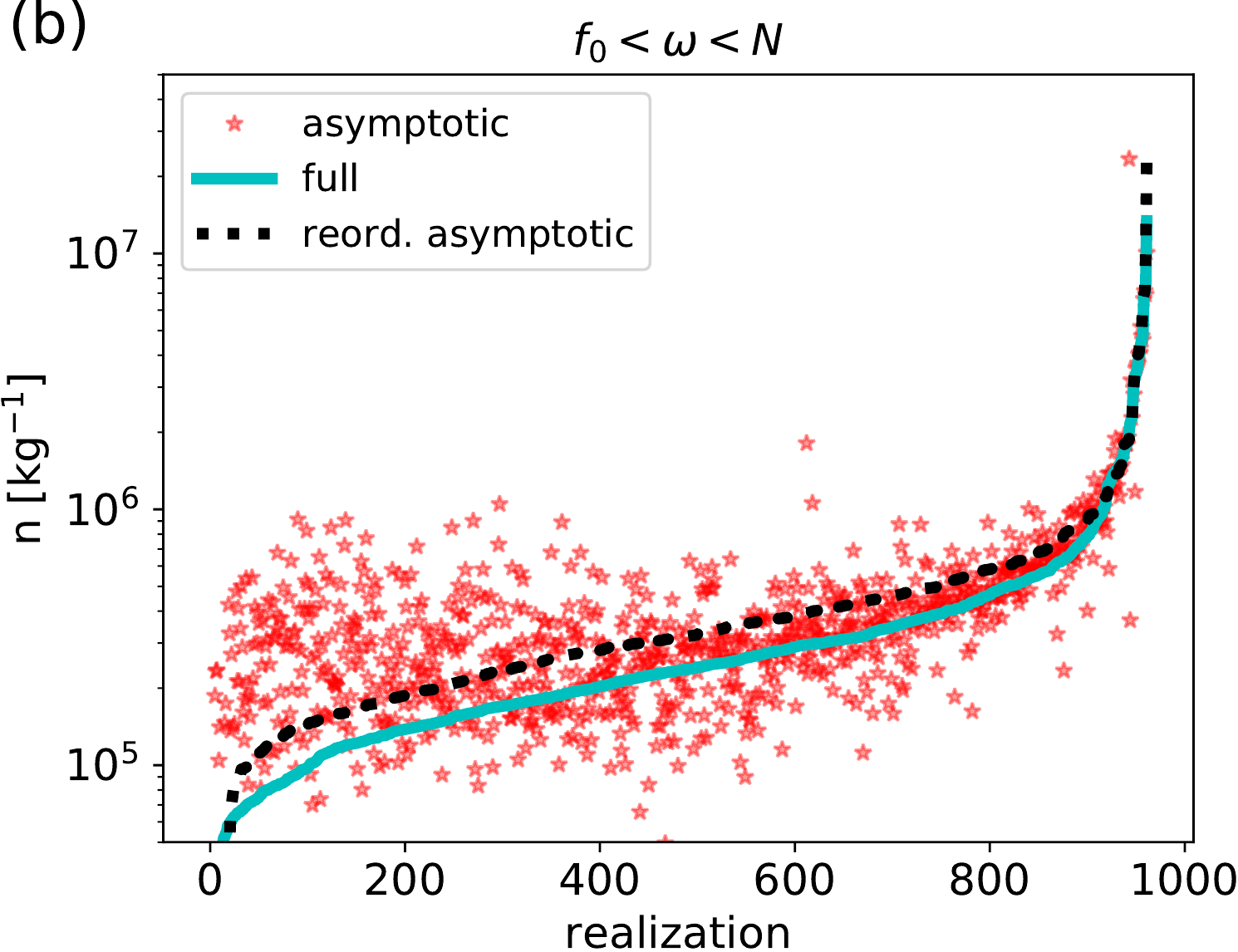} 
  \caption{\revision{Nucleated number concentration, $n$, for $10^3$
      different realizations of a superposition of GW with
      frequencies: $ 0.5T^{-1}_w < \omega < 2 T^{-1}_w$ (a) and
      $f_0 < \omega < N$ (b). The cyan line shows in ascending order
      the results from the full model. The corresponding asymptotic
      solutions are shown as red stars. The dashed black line displays
      the same asymptotic solutions but after sorting them in
      ascending order.}}
\end{figure}

\revision{In this section we consider the case when a GW spectrum is
  forcing the ice physics. The GW forcing is constructed by
  considering $N_{GW}$ waves with vertical wind amplitudes sampled
  from a white frequency spectrum in accordance with observations
  \citep{podglajen_lagrangian_2016}. The wave amplitudes $\hat w$ are
  rescaled such that the total GW momentum flux, $\rho u' w' $, equals
  5 mPa at 8 km altitude. This value for the momentum flux is within
  the range given by observational and modeling studies
  \citep{hertzog_intermittency_2012, kim_toward_2021,
    corcos_observation_2021}. For high-frequency GW (with
  $T_w \sim N$) in the tropopause region the ratio
\begin{align}
  \frac{T_d}{T_w} = \Order{\eps^{-1}}
\end{align}
is more appropriate, compare the last equation with \eq{dist_limit_t}.
In this case, one obtains from \eq{transf_dot_pi}
\begin{align}
  \label{transf_dot_pi_high_freq}
  \DD{\pi}{t_d} = \left(\eps^{1+\alpha}\DD{\pi'}{t_w}
  + \eps w\dd{\bar \pi}{z_s} \right)\, ,
\end{align}
implying that the tendency of $\pi'$ is of the same order as the
vertical advection term for $\alpha=0$. It was verified that the
magnitude of the wave amplitudes allows to neglect the nonlinear
advection term and the term $\pp{\pi'}{t}$ will have the dominant
contribution in the material derivative of $\pi'$. In order to account
for effects due the high-frequency waves, we will include the latter
term into the GW forcing term. We will also include a next order
correction term by keeping the $c_p/R$ term in equation \eq{S},
since we found that this improves the results for the general case of
superposition of GW.  We found from simulations that the magnitudes of
the waves are too small to trigger nucleation, because of this we include a
constant vertical updraft $w_{00}$=2 cm s$^{-1}$ on which the GW are
superimposed. With the assumptions above, the forcing \eq{def_a} is
generalized to}
\revision{
\begin{align}
  \label{def_a_high_freq}
  F^*(t_w) = 
  \left(L^* - \eps \frac{c_p}{R} \right)
  \left\{
  \sum \limits_j^{N_{GW}} \left[   \frac{R}{c_p}
  |\tilde w_j^{(0)}| \cos(\omega_j t_w + \phi_j)
  + \omega_j |\tilde \pi_j^{(0)}| \sin(\omega_j t_w + \phi_j)
  \right]
  + \frac{R}{c_p} w_{00}   
  \right\}
\end{align}
}
\revision{With the new forcing \eq{def_a_high_freq} we compute the ice
  crystal number concentration using the asymptotic solution from
  Section~\ref{derivation}\ref{summary_asy}.  We perform $10^3$
  realizations, each forced by a superposition of $10$ waves with
  random frequencies uniformly distributed within the range
  $\omega_{min} < \omega_j < \omega_{max}$. A random wave phase
  increment
  $\delta \phi_j = \phi_j - \mathbf{x}(t_*) \cdot \mathbf{k}_j$ is
  uniformly distributed within $[0, 2 \pi]$. The vertical wavelength
  of all waves is set to 1 km and the initial values $n=q=0$ and
  $S=1.4$ are used in the simulations.

  The results are summarized in Fig.~\ref{ensemble}a for the case
  where $\omega_j$ is drawn from a narrow range around the frequency
  $T_w^{-1} = 2\times 10^{-3}$ s$^{-1}$ for which the asymptotic
  analysis was performed. The asymptotic solution captures the ice
  crystal number concentration, $n$, from the full model for the
  majority of realizations. The accuracy increases if large values of
  $n$ are nucleated. Further, by reordering $n$ from the asymptotic
  solution in ascending order (the dashed black line in
  Fig.~\ref{ensemble}a), we see that the distribution produced by the
  asymptotic model is identical with the one from the full model (cyan
  line). This is an important result for the application of the
  present asymptotic solutions in climate models. Since the phase of
  the waves is typically unknown, any parameterization will not be
  able to reproduce the exact nucleation time $t_0$. However, on
  average our model produces a distribution of $n$ matching the one
  from the full model.

  Next, we consider the full frequency range of GW, the corresponding
  results are summarized in Fig.~\ref{ensemble}b. It has to be
  stressed, that our asymptotic analysis is only valid for the
  frequency range around $T_w^{-1}$. However, the asymptotic model can
  still capture many of the nucleation events of the full
  model. Again, this is particularly valid for the largest value
  $10^6 < n < 10^7$, which will probably be the more relevant. For
  smaller values of $n$ the quality of prediction
  deteriorates. Reordering of the asymptotic values reveals a small
  positive bias in the distribution produced by the asymptotic
  model. We suppose that this bias might be due to small amplitudes of
  the GW, inconsistent with our scaling where $\hat w \sim W_c$ was
  assumed, and the omission of the low-frequency GW in the asymptotic
  analysis.}

\section{Conclusions}
\label{conclusions}
We present an asymptotic approach allowing to identify a reduced model
for the self-consistent description of ice physics forced by a
superposition of GW including the effect of diffusional growth and
homogeneous nucleation of ice crystals. Further, using matched
asymptotic techniques analytical solutions are constructed, involving
a novel parameterization \eq{n_final} for the ice crystal number
concentration $n$. \revision{The latter has as input parameters the
  wave amplitudes and phases, and the time of the nucleation event.
  It allows the derivation of upper bound for the nucleated $n$, as
  well as, a threshold for the initial $n$ which will inhibit
  nucleation.  The numerical simulations with a Lagrangian parcel
  model show that the parameterization reproduces nucleation events
  triggered by a monochromatic GW for a variety of initial
  conditions. Further, in the case of superposition of GW within the
  mid-frequency range the parameterization generates a distribution of
  $n$ matching the one of the full model. By extending the
  parameterization to high-frequency GW it is shown that the
  asymptotic solution produces distribution similar to the one of the
  full model even if the complete GW frequency spectrum is used as
  forcing.  The results presented here demonstrate the potential of
  our approach for constructing improved cirrus schemes in climate
  models with realistic GW variability as simulated with transient GW
  parameterizations \citep{boloni_toward_2021, kim_toward_2021}.}
  
When comparing the treatment of the ice physics in our approach with
the one from BS19, we observe different scaling in the nucleation
term: in the latter work $J \sim B \sim \eps^{-1}$ is used, whereas
here we apply $J \sim \eps^{-1}, B \sim \eps^{-2}$. Nevertheless, our
parameterization \eq{n_final} is equivalent to the closure of BS19 for
constant updraft velocity if the velocity there is replaced by the GW
vertical velocity at the nucleation time $t_0$. This is not
surprising, since the GW nearly does not vary on the fast nucleation
time scale. The correspondence of the two parameterizations becomes
more clear if one takes into account that in BS19
$\eps=\Order{10^{-2}}$ and here $\eps=\Order{10^{-1}}$, implying the
same magnitude of the nucleation exponent $B$ under the different
scalings. As shown by \cite{spichtinger_impact_2022} the exact value
of the nucleation rate $J$ is not crucial as long as it is
sufficiently large. \revision{Still, we have to stress that the
  present approach generalizes the framework of BS19 to include wave
  dynamics and the consistency between the two parameterizations only
  supports our results. In addition, we derive a novel
  parameterization for the variable mean mass model, see
  \eq{alg_eq_n_post}, a threshold for nucleation inhibition,
  $N^{max}_{pre}$ from \eq{n_max_pre}, and an high-frequency
  correction to the parameterization, \eq{def_a_high_freq}. }


The present asymptotic solutions are applicable mainly to the
mid-frequency GW in the troposphere and tropopause region, as well as,
to high-frequency GW in the troposphere.  For high-frequency GW in the
tropopause region the ratio between the time scale of the diffusional
growth and of the wave is given by $T_d/T_{w}\sim \eps^{-1}$. \revision{The
simulations from Sec.~\ref{superposition} shows large numbers of $n$ if
the high-frequency GW are included. This suggests a new regime
dominated by the GW forcing term and we propose some asymptotic
corrections to account for it.} We expect that this regime corresponds
to the temperature-limit events studied in \cite{dinh_effect_2016}.
For low-frequency GW the scaling $T_d/T_{w}\sim \eps$ is
appropriate. In this case the GW forcing term becomes by a factor of
$\eps$ weaker, when compared to the depositional growth term. This
regime is relevant for low updraft velocities and will be considered
in an upcoming study.

\revision{Our asymptotic analysis assumes a reference number
  concentration, $n_c$. However, the results from
  Sec.~\ref{superposition} suggest that the resulting asymptotic model
  is valid for a wider range of $n$. If regimes with other values of
  $n_c$ are of interest, the present asymptotic framework can be
  adapted for the systematic investigation of these too.

  The models presented here predict for the particular GW forcings
  values of $n$ which are within or at the upper range of
  observations, e.g., see Fig. 8 from
  \cite{kramer_microphysics_2020}. However, a direct comparison with
  observational data is hampered for two main reasons. First, most of
  the measurements lack information on the wave properties, e.g., wave
  amplitude and frequency, so the GW forcing cannot be
  determined. Second, nucleation takes place at very fast time scale
  and within a confined spatial region. Therefore, the vast majority
  of measurements of ice crystals are taken probably after the
  nucleation event happened. However, at later stages of the ice cloud
  life cycle, other processes such as sedimentation are determining
  the microphysical properties. The latter lead to smearing of the
  clear nucleation signature and to significantly smaller $n$ values
  \citep[see, e.g.,][]{spichtinger_modelling_2009}; this effect is
  enhanced if ice crystals fall into subsaturated air and thus
  evaporate. This might explain, why high number densities are quite
  rarely observed \citep[see, e.g.,][] {kramer_ice_2009,
    kramer_microphysics_2020}.}

In the present regime the magnitude of the sedimentation effects is
determined by the sedimentation time scale
$T_{sed}=H_c/c_q m_c^{2/3}$. Substituting the reference quantities
gives $T_{sed} \sim\eps^{-2} T_d \sim 11$ h, implying that at leading
order sedimentation is negligible compared to the diffusional growth
term. Note however, that $T_{sed}$ will decrease if regimes with
larger ice crystal mass, $m_c$, or smaller vertical scales, $H_c$, are
of interest. As shown in \cite{podglajen_impact_2018}, sedimentation
modulated by GW forcing produces localization effects in cirrus.

In the present study only cirrus formed by homogeneous nucleation are
considered, since this is the dominant formation mechanism in the cold
temperature regime with strong updraft velocities
\citep[e.g.,][]{heymsfield_homogeneous_1993}. Still, heterogeneous
nucleation can considerably alter cirrus formation \citep[see,
e.g.,][]{gierens2003,spichtinger_impact_2010}; however, the important
feature is, also in case of competing nucleation pathways, the
occurrence of pre-existing ice crystals, as in our investigations.  In
addition, turbulence due to GW breaking is another source of GW
generated variability omitted in the present study \revision{\citep[e.g.,][]{atlas_aircraft_2023}}.


%
\acknowledgments UA ans PS thank the German Research Foundation (DFG)
for partial support through the research unit Multiscale Dynamics of
Gravity Waves (MS-GWaves), TRR 301 – Project-ID 428312742 ``TPChange''
Projects B06 ``Impact of small-scale dynamics on UTLS transport and
mixing'' and B07 ``Impact of cirrus clouds on tropopause structure'', and
through Grants AC 71/8-2, AC 71/8-2, and AC 71/12-2. UA acknowledges
partial support through Grant CRC 181 ``Energy transfers in Atmosphere
an Ocean'', Project Number 274762653, Projects W01 ``Gravity-wave
parameterization for the atmosphere'' and S02 ``Improved
Parameterizations and Numerics in Climate Models.''

\datastatement
The Python script used to generate all figures in the paper is available upon
request.


\appendix[A]
\appendixtitle{Time evolution of $S$}
\label{appendix_s}
\revision{
Here we derive from \eq{q_v} an evolution equation for the saturation
ratio $S$. Since, $q_{v,c}=\eps_0 p_{si,c}/p_{00}$ is used for the
scaling of $q_v$, the definition of $S$ expressed using nondimensional
variables reads
\begin{align}
  \label{def_s}
  S=\frac{q_v p}{p_{si}}\, .
\end{align}
Applying $\dd{}{t_d}$ to \eq{def_s} yields
\begin{align}
  \DD{S}{t_d} & 
  =\frac{p}{p_{si}}\DD{q_v}{t_d} + \frac{S}{p}\DD{p}{t_d} -
  \frac{S}{p_{si}}\DD{p_{si}}{t_d}  \\ 
  \label{line_2}
  &=\frac{p}{p_{si}}\DD{q_v}{t_d} + \frac{c_p S}{R
    \pi}\DD{\pi}{t_d} - \frac{SL^*}{\eps T^2}\DD{T}{t_d} \\ 
  \label{line_3}
  &=\frac{p}{p_{si}}\DD{q_v}{t_d} + \frac{c_p S}{R
    \pi}\DD{\pi}{t_d} 
  -  \frac{SL^*}{\eps T^2}\left(\pi \underbrace{\DD{\theta}{t_d}}_{=0} +
  \theta \DD{\pi}{t_d} \right) \\ 
  \label{line_4}
  &=\frac{p}{p_{si}}\DD{q_v}{t_d} -
  \frac{S}{\pi}\DD{\pi}{t_d}\left(\frac{L^*}{\eps T} - 
  \frac{c_p}{R}\right) 
\end{align}
where \eq{cla_cla} was used to obtain \eq{line_2}, \eq{nond_dot_theta}
in \eq{line_3} and the definition of potential temperature
$\theta = T/\pi$ for the last equation \eq{line_4}. With this one
obtains \eq{S}.}

\appendix[B]
\label{app:gw_omega}
\appendixtitle{GW dispersion relation and polarization relations}

\revision{ With $\mathbf{v} = \mathbf{v}^{(0)} + \Order{\eps}$ the
  leading order continuity equation reads
\begin{align}
  \nabla \cdot \mathbf{v}^{(0)} = 0.
\end{align}
Using the wave ansatz for $\mathbf{v}^{(0)}$, this gives the 
solenoidality condition
\begin{align}
 \label{solenoid_cond} 
\mathbf{k} \cdot \tilde {\mathbf{ v}}^{(0)} = 0\, ,
\end{align}
implying that the wavevector $\mathbf{k}$ and
$\tilde {\mathbf{ v}}^{(0)}$ are orthogonal. The latter property will
be used to eliminate nonlinear advection terms in the equations.

From the leading order vertical momentum equation we obtain
hydrostatic balance between $\bar \pi^{(0)}$ and $\bar \theta^{(0)}$
\begin{align}
\label{pi_zero}
  \dd{\bar \pi^{(0)}}{z_s} = - \frac{R}{c_p \bar \theta^{(0)}}\, .
\end{align}
Similarly, the next order vertical momentum balance reads
\begin{align}
\dd{\bar \pi^{(1)}}{z_s} = \frac{R\bar \theta^{(1)}}{c_p \bar \theta^{(0)^2}}\, ,
\end{align}
where we have used \eq{pi_zero}. Note, that in the case
$\alpha=\beta=0$ there is no advection term appearing in the latter
equation due to \eq{solenoid_cond}.

The projection of the leading order equations onto the GW field reads
\begin{align}
\label{wave_1}
 \pp{\vu^{(0)}}{t_w}  &= - \frac{c_p \bar \theta^{(0)}}{R} \nabla_{h} \pi^{(2+\alpha)} \\
(1-\beta) \pp{w^{(0)}}{t_w}
&= - \frac{c_p \bar \theta^{(0)}}{R}\pp{\pi^{(2+\alpha)}}{z_w} 
+ \frac{\theta^{(1+\alpha)}}{\bar \theta^{(0)}} \\
\nabla_w \cdot \vvb^{(0)} &= 0\, ,\\
\label{wave_4}
 \pp{\theta^{(1+\alpha)}}{t_w} + w^{(0)} \dd{\bar \theta^{(\alpha)}}{z_s} &= 0\, ,
\end{align}
where again \eq{solenoid_cond} was utilized. Inserting in
eqs. \eq{wave_1}-\eq{wave_4} a wave ansatz for the solution, one
obtains the following system of linear equations for the wave
amplitudes
\begin{align}
  \label{hom_lgs}
  \bm{\mathsf{M}} \mathbf{z} = 0
\end{align}
where
\begin{align}
\bm{\mathsf{M}} = 
\left(
  \begin{array}[h]{c c c c c}
   - i \omega & 0 & 0 & 0 & ik \\
     0 & -i \omega & 0 & 0 & il \\
     0 & 0 & -i (1-\beta)\omega & -\bar N & im \\
     0 & 0 & \bar N & -i \omega & 0 \\
     ik & il & im & 0 & 0        
  \end{array}
\right)
\end{align}
with $\mathbf{z} = (\tilde u^{(0)}, \tilde v^{(0)},\tilde w^{(0)},\tilde b^{(1+\alpha)}/ \bar N, \frac{c_p}{R}\bar \theta^{(0)} \tilde \pi^{(2+\alpha)})$ and
\begin{align}
  \label{def_N_bar}
  \bar N^2&=\frac{1}{\bar \theta^{(0)}}\dd{\bar \theta^{(\alpha)}}{z} \\
   \label{def_b}
  \tilde b^{(1+\alpha)} &= \frac{\tilde \theta^{(1+\alpha)}}{\bar \theta^{(0)}}\, .
\end{align}
Looking for non-trivial solutions of \eq{hom_lgs}, one derives the
dispersion relation \eq{omega_mid-freq} and the polarization relations
\eq{polarization} for the GW amplitudes.}

\appendix[C]
\revision{
  \appendixtitle{Evolution equation for $n$ in the nucleation regime}
\label{evol_n}  
  Here, the equation for $n$ is derived: first \eq{n_tau_1ode_ord} is
  differentiated with respect to $\tau$ and after this the exponential
  function is replaced using \eq{n_tau_1ode_ord} giving
\begin{align}
  \label{n_tau2}
  \ddn{n}{\tau}{2} &= \frac{B^*}{\eps^2} \dd{S}{\tau}
                     \dd{n}{\tau} + \Order{\eps^2} \, .
\end{align}
Inserting \eq{S_tau_1ode_ord} yields
\begin{align}
  \ddn{n}{\tau}{2} &=  B^* \dd{n}{\tau} \left[ - D^* (S-1) T n 
 - \frac{L^* S w}{\bar \pi} \dd{\bar \pi}{z_s}
 \right] + \Order{\eps}\, .
\end{align}
%
The evaluation of the leading order equation takes the form
\begin{align}
  \ddn{n^{(0)}}{\tau}{2} &=  B^* \dd{n^{(0)}}{\tau}\left[ - D^* (S^{(0)}-1) \bar T^{(0)} n^{(0)} -
   \frac{L^* S^{(0)} w^{(0)}}{\bar \pi^{(0)}} \dd{\bar \pi^{(0)}}{z_s}  \right]\, .
\end{align}
After substituting \eq{bar_pT_z00}, \eq{w0_cos}, \eq{pi_zero} and
\eq{def_a} in the last equation one yields
\begin{align}
  \label{n_tau_2ode}
  \ddn{n^{(0)}}{\tau}{2} &=  B^* \dd{n^{(0)}}{\tau}\left[
                           - D^* (S^{(0)}-1) n^{(0)} +
                           \revision{F^*(t_0) S^{(0)} }\right]\, ,
\end{align}
where the expansion:
$F^*(t_w) = F^*(t_0) + \Order{\eps^2}$ was used.
Taking into account that $S^{(0)}=S_c$, equation \eq{n_tau_2ode} can be written as
\begin{align}
  \label{n_dd}
    &\dd{}{\tau} \left(\dd{n^{(0)}}{\tau} + \delta (n^{(0)})^2 - \gamma n^{(0)}
      \right) = 0
\end{align}
with the definitions introduced in \eq{delta}, \eq{gamma}. Integrating \eq{n_dd}
over time finally gives \eq{def_m}.
}

\appendix[D]
  \appendixtitle{Composite solution}
\revision{
Despite the fact that $N_{post}$ is determined from \eq{n_final}, the
integration constant $n_0$ entering \eq{n_tau} through \eq{def_C} is
still unknown. It will, however, not effect the value of $N_{post}$.
Moreover, since in the present asymptotic analysis $t_0$ can be found
up to some higher order corrections, $n_0$ is undetermined. To see
this we introduce another constant $\tau_0$ defined as
$C= e^{-\sigma \tau_0}$, we can write \eq{n_tau} as
\begin{align}
    \label{n_tau_0}
  n^{(0)}(\tau)  = \frac{n_s + n_e e^{\frac{\sigma}{\eps^2} (t-t_0 -
      \eps^2 \tau_0)}}{1 + e^{\frac{\sigma}{\eps^2} (t-t_0 - \eps^2
      \tau_0)}} 
\end{align}
Note, that from \eq{value_t0} $t_0$ is determined up to $\Order{\eps}$
corrections, which will result in modifications of the constant $n_0$.
One way of setting the value for $n_0$ is by requiring that at
$\tau=0$ the saturation ratio $S$ should reach a maximum. At the end
of the pre-nucleation we have $\dot S >0$, on the other hand at the
beginning of the post-nucleation $\dot S <0$, thus $S$ should have a
maximum within the nucleation regime. We define $n_0$ by requiring
that in \eq{S_tau_1ode_ord} $\dot S=0$ at time $\tau=0$ up to
$\Order{\eps^3}$ corrections. After using \eq{bar_pT_z00}, \eq{w0_cos}, \eq{pi_zero} \eq{def_a} and \eq{s_eq_sc} this implies
\revision{
\begin{align}
  \label{n_0}
  n_0
  = \frac{F^*(t_0) S_c}{D^* (S_c -1)} \implies C = 1 \, .
\end{align}
}
The requirement of having a maximum in the saturation ratio upon the
nucleation is physically meaningful and may be interpreted as the
defining feature of a nucleation event.

It remains to construct the composite solution valid in all three
regimes. For the number concentration the nucleation regime represents
an interior layer \citep[e.g.][]{holmes}, enclosed by the outer layers
of the pre-nucleation and post-nucleation regime. In this case the
composite solution reads
\begin{align}
\label{n_comps_def}
  n(t_w) = n^{(0)}_{pre}(t_w) + n^{(0)}_{nuc}(\tau) + n^{(0)}_{post}(t_w) - n^{(0)}_{nuc}(-\infty) - n^{(0)}_{nuc}(\infty) + \Order{\eps}\, .   
\end{align}
Substituting \eq{n_pre_post}, \eq{n_tau}, \eq{n_tau_minfty} and \eq{n_tau_infty}
in \eq{n_comps_def} gives for the number concentration
\begin{align}
  \label{n_comps_ndim}
  n(t_w) &= \frac{n_s + n_e e^{\frac{\sigma}{\eps^2}(t_w-t_0)}}{1 + e^{\frac{\sigma}{\eps^2}(t_w-t_0)}} + \Order{\eps}\, .
\end{align}
Since the time derivative of the saturation ratio has a jump from the
pre-nucleation to the post-nucleation value, the nucleation regime represents a corner layer for $S$ \citep{holmes}. For such a layer two cases depending on
the sign of $t-t_0$ has to be considered when constructing
the composite solution
\begin{align}
  \label{S_comps_def}
  S(t_w) =
  \begin{cases}
  S^{(0)}_{pre}(t_w) + S^{(0)}_{nuc}(\tau) - S^{(0)}_{nuc}(-\infty)\qquad &\text{ for } t_w \le t_0 \\
 S^{(0)}_{post}(t_w) + S^{(0)}_{nuc}(\tau) - S^{(0)}_{nuc}(\infty)\qquad    &\text{ for } t_w >t_0
  \end{cases}+ \Order{\eps}\, .
\end{align}
Substituting \eq{S_pre_post}, \eq{s_eq_sc} in \eq{S_comps_def}, the solution for $S$ takes the form
\begin{align}
  \label{S_comps_ndim}
  S(t_w) =
  \begin{cases}
    S_*S_h(t_w,t_*) + \int \limits_{t_*}^{t_w} dt' D^* n_s
    S_h(t_w,t') \qquad &\text{ for } t_w \le t_0 \\
    S_c S_h(t_w,t_0) + \int \limits_{t_0}^{t_w} dt' D^* n_e
    S_h(t_w,t') \qquad &\text{ for } t_w>t_0
  \end{cases}+ \Order{\eps}\, ,
\end{align}
with $S_h$ defined in \eq{S_h}.  
}  
  \appendix[E]
\revision{
  \appendixtitle{Description Lagrangian parcel model}
The Lagrangian parcel model describes the evolution of $n, q$ and
$q_v$ in an air parcel oscillating in the $x-z$ plane under the
gravity wave forcing. It solves \eq{dot_n_non}-\eq{q_v} with
sedimentation switched off, i.e., $S^*_n=S^*_q=0$.  The parcel
position vector, $\mathbf{x}(t)=(x, z)^T$, is determined from
\begin{align}
  \label{trajectory}
  \frac{d\mathbf{x}}{dt} = \mathbf{v}, 
\end{align}
where the velocity field is given in general by a superposition of GWs
with a possibility to include a constant updraft velocity, $w_{00}$,
\begin{align}
  \mathbf{v} =
  \sum_j^{N_{GW}}  \mathbf{\tilde v}_j \cos 
  \left(\omega_j t  + k_j x + m_j z  + \delta \phi_j \right) 
  + w_{00} \mathbf{e}_z\, . 
\end{align}
Each frequency $\omega_j$ and amplitude $\mathbf{\tilde v}_j$ is
satisfying the general inertial GW dispersion relation and
polarization relation \citep[e.g.][]{achatz_atmospheric_2022} with
$N = 0.02 $ s$^{-1}$ for the the tropopause region. A vertical
wavelength of 1 km is assumed, comparable to the value of 3 km used in
the study of \cite{corcos_simple_2023}.

From \eq{trajectory} with initial condition $z=z_{00}$ and $x=x_{00}$
the parcel position is found. With this the wave fluctuations of Exner
pressure, $\pi'$, and of potential temperature, $\theta'$, are
determined from the corresponding polarization relations. To those
fluctuations one has to add the stationary contributions $\bar \pi$
and $\bar \theta$ from the reference atmosphere in order to determine
the full fields. It is assumed that in the vicinity
of $z_{00}$ the reference atmosphere can be represented by an
isothermal temperature profile $\bar T(z)=T_{00}$ with corresponding
pressure $\bar p(z) = p_{00} e^{-\frac{z-z_{00}}{H_p}}$ (all in
dimensional form). From those the Exner pressure, $\bar \pi$, and
pontential temperature, $\bar \theta$, of the reference atmosphere are
calculated using the definitions $T = \pi \theta$ and
$\pi = (p/p_{00})^{R/c_p}$. By adding all together one obtains the
total fields $\pi$ and $\theta$, or equivalently $p$ and
$T$. Finally, the saturation pressure can be determined from
\eq{p_sat_ice}. In practice, equations \eq{dot_n_non}-\eq{q_v} and
\eq{polarization} are simultaneously integrated numerically and we use
the above mentioned procedure to find $p$, $p_{si}$ and $T$ at each
time step.
}

\bibliographystyle{ametsocV6}
\bibliography{references}


\end{document}